\documentclass[aps,prd,showpacs,eqsecnum,twocolumn,superscriptaddress,a4paper,nofootinbib]
{revtex4-1}
\usepackage{amsmath,amssymb,graphicx,color}
\usepackage{bm}
\usepackage{epsf}

\begin{document}

\title{Dynamical mass ejection from the merger of asymmetric binary
  neutron stars: \\ Radiation-hydrodynamics study in general relativity}

\author{Yuichiro Sekiguchi} 
\affiliation{Department of Physics, Toho
  University, Funabashi, Chiba 274-8510, Japan}

\author{Kenta Kiuchi}
\affiliation{Yukawa Institute for Theoretical Physics, 
Kyoto University, Kyoto, 606-8502, Japan~} 

\author{Koutarou Kyutoku} 
\affiliation{Interdisciplinary Theoretical
Science (iTHES) Research Group, RIKEN, Wako, Saitama 351-0198, Japan}

\author{Masaru Shibata}
\affiliation{Yukawa Institute for Theoretical Physics, 
Kyoto University, Kyoto, 606-8502, Japan~} 

\author{Keisuke Taniguchi}
\affiliation{Department of Physics, University of the Ryukyus,
Nishihara, Okinawa 903-0213, Japan}

\date{\today}

%%%%%%%%%%%%%%%%
\begin{abstract}
%%%%%%%%%%%%%%%% 

We perform neutrino radiation-hydrodynamics simulations for the merger
of asymmetric binary neutron stars in numerical relativity.  Neutron
stars are modeled by soft and moderately stiff finite-temperature
equations of state (EOS). We find that the properties of the dynamical
ejecta such as the total mass, neutron richness profile, and specific
entropy profile depend on the mass ratio of the binary systems for a
given EOS in a unique manner.  For the soft EOS (SFHo), the total
ejecta mass depends weakly on the mass ratio, but the average of
electron number per baryon ($Y_e$) and specific entropy ($s$) of the
ejecta decreases significantly with the increase of the degree of mass
asymmetry. For the stiff EOS (DD2), with the increase of the mass
asymmetry degree, the total ejecta mass significantly increases while
the average of $Y_e$ and $s$ moderately decreases.  We find again that
only for the soft EOS (SFHo), the total ejecta mass exceeds
$0.01M_\odot$ irrespective of the mass ratio chosen in this paper. The
ejecta have a variety of electron number per baryon with its average
approximately between $Y_e \sim 0.2$ and $\sim 0.3$ irrespective of
the EOS employed, which is well-suited for the production of the
r-process heavy elements (second and third peaks), although its
averaged value decreases with the increase of the degree of mass
asymmetry.

%%%%%%%%%%%%%%
\end{abstract}
%%%%%%%%%%%%%%

\pacs{04.25.D-, 04.30.-w, 04.40.Dg}

\maketitle

%%%%%%%%%%%%%%%%
% Introduction %
%%%%%%%%%%%%%%%% 

\section{Introduction}

The merger of binary neutron stars is one of the most promising
sources of gravitational waves for ground-based advanced detectors,
such as advanced LIGO, advanced VIRGO, and
KAGRA~\cite{Detectors}. Among them, advanced LIGO already started the
first observational run and has achieved the first direct detection of
gravitational waves, which were emitted from a binary-black-hole
merger~\cite{abbott16}.  We should expect that these
gravitational-wave detectors will also detect the signals of
gravitational waves from binary-neutron-star mergers in a few years,
because the latest statistical studies suggest that these
gravitational-wave detectors will observe gravitational waves from
merger events as frequently as $\sim 1$--$100$/yr if the designed
sensitivity is achieved~\cite{RateLIGO,dominik,Kim}.

Binary-neutron-star mergers are also attracting attention as one of
the major nucleosynthesis sites of heavy elements produced by the
rapid neutron capture process (r-process)~\cite{Lattimer}, because a
significant fraction of the neutron-rich matter is likely to be
ejected during the merger (see Ref.~\cite{Rosswog} for the pioneering
work). Associated with the production of the neutron-rich heavy
elements in the matter ejected during the merger, a strong
electromagnetic emission could be accompanied by the radioactive decay
of the r-process heavy elements~\cite{Li:1998bw,KN1,TH}. This will be
an electromagnetic counterpart of gravitational waves from
binary-neutron-star mergers and its detection could be used to verify
the binary-neutron-star-merger scenario for the r-process
nucleosynthesis.  This hypothesis is encouraged in particular by the
observation of an infrared transient event associated with a
short-hard gamma-ray burst, GRB\,130603B~\cite{GRB130603B}.  These
facts strongly encourage the community of gravitational-wave astronomy
to theoretically explore the mass ejection mechanisms, the r-process
nucleosynthesis in the ejecta, and associated electromagnetic emission
in the mergers of binary neutron stars.

For the quantitative study of these topics, we have to clarify the
merger process, subsequent mass ejection, physical condition of the
ejecta, nucleosynthesis and subsequent decay of the heavy elements in
the ejecta, and electromagnetic emission from the ejecta. For these
issues, a numerical-relativity simulation, taking into account the
detailed microphysical processes and neutrino radiation transfer, is
the unique approach. In our previous paper~\cite{Sekig2015}, we
reported our first numerical-relativity results for these issues
focusing only on the equal-mass binaries. We found that the total mass
of the dynamically ejected matter during the merger depends strongly
on the equations of state (EOS) we employed, while the ejecta
components have a wide variety of electron number per baryon
(denoted by $Y_e$) between $\approx 0.05$ and $\approx 0.5$ irrespective of
the EOS employed (see also Refs.~\cite{Palenzuela2015,foucart2015,ber2015,radice2016}).
The broad $Y_e$ distribution is well-suited for explaining 
the abundance patterns for the r-process heavy elements with 
the mass number larger than $\sim 90$ in the solar system 
and ultra metal-poor stars~\cite{Wanajo}.  

In this article, we extend our previous study
focusing on the merger of asymmetric binary neutron stars: We will
report our latest numerical results for unequal-mass binary systems of
typical neutron-star mass (between 1.25 and $1.45M_\odot$) for a soft
(SFHo) EOS~\cite{SFHo} and a moderately stiff (DD2) EOS~\cite{DD2}. We
will show that the physical properties of the merger ejecta depend on
the degree of the mass asymmetry of the system: The ejecta mass varies
with the mass ratio for a fixed value of the binary total mass, and the
averaged value of $Y_e$ decreases with the increase of the mass
asymmetry degree, although $Y_e$ is always broadly distributed
irrespective of the mass ratio.

% which is a well-suited qualitative
%property for explaining the abundance patterns of r-process elements
%in the solar system and ultra metal-poor stars.

The paper is organized as follows.  In Sec.~II, we briefly review the
formulation and numerical schemes employed in our numerical-relativity
simulation, and also summarize the EOS we employ.  In Sec.~III, we
present numerical results focusing on the dynamical mass ejection and
properties of the merger remnants.  Section~IV is devoted to a
summary.  Throughout this paper, $c$ and $G$ denote the speed of light
and the gravitational constant, respectively. 

%%%%%%%%%%%%%%%%%%%%%%%%%%%%%%%%
% Method and initial condition %
%%%%%%%%%%%%%%%%%%%%%%%%%%%%%%%%

\section{Method, EOS, initial models, and grid setup.} 

We solve Einstein's equation by a
puncture-Baumgarte-Shapiro-Shibata-Nakamura formalism as
before~\cite{BSSN,Sekig,Sekig2015}.  The fourth-order
finite-differencing scheme is applied to discretize the field
equations except for the advection terms for which the lop-sided
scheme is employed.  The radiation hydrodynamics equations are solved
in the same manner as in Ref.~\cite{Sekig2015}: Neutrino radiation
transfer is computed in a leakage scheme~\cite{Sekig2012}
interpolating Thorne's moment formalism with a closure relation for a
free-streaming component~\cite{Kip,MKSY}. For neutrino heating, which
could induce a neutrino-driven wind from the merger remnant,
absorption on free nucleons is taken into account.

%% EOS

We employ a soft (SFHo)~\cite{SFHo} and a moderately stiff
(DD2)~\cite{DD2} EOS for the nuclear-matter EOS, which have been
derived recently by Hempel and his collaborators.  For these EOS, the
predicted maximum mass for spherical neutron stars is $2.06M_\odot$
and $2.42M_\odot$, respectively, and larger than the largest
accurately-measured mass of neutron stars, $\approx
2.0M_\odot$~\cite{Demorest:2010bx}. The radius of neutron stars with
mass $1.35M_\odot$ for them is $R_{1.35}=11.9$~km (SFHo EOS) and
$13.2$~km (DD2 EOS), respectively. These radii depend very weakly on
the mass as long as it is in a typical neutron-star mass range between
1.2 and $1.5M_\odot$. Thus, we refer to an EOS with $R_{1.35} \leq
12$~km like SFHo EOS as soft EOS. The stellar radius plays a key role for
determining the merger remnant and the properties of the dynamical
ejecta as we already described in our previous paper~\cite{Sekig2015}.

%% Numerical set up: !!!! WE NEED UPDATA

\begin{table*}[t]
\centering
\caption{\label{tab1} The parameters and the results of the models
  employed in this study.  $m_1$ and $m_2$: each mass of binary in
  isolation. $q$: mass ratio defined by $m_2/m_1 (\leq 1)$. $\Delta
  x_{9}$: the grid spacing in the finest refinement level. $N$: the
  grid number in one positive direction for each refinement level.
  $M_{\rm ej}$, $\langle Y_{e} \rangle$, and $V_{\rm ej}$ denote the
  dynamical ejecta mass, the averaged value of $Y_{e}$, and ejecta
  velocity measured at 30\,ms after the onset of the merger.  $M_{\rm
    BH}$ and $a_{\rm BH}$ are the mass and dimensionless spin
  parameter of the remnant black holes, and $M_{\rm torus}$ is the mass
  of tori surrounding the remnant black holes for the SFHo models.
  These values are also measured at 30\,ms after the onset of the
  merger.  Model name follows the EOS, each mass $m_2$ and $m_1$, and
  grid resolution.  The equal-mass data are taken from
  Ref.~\cite{Sekig2015}.  }
%\begin{minipage}{140mm}
\begin{tabular}{cccccccccccc}
\hline\hline
Model & $(m_1, m_2)$ & $q=m_2/m_1 $& 
$\Delta x_{9}$\,(m) & ~~$N$~~ & $M_{\rm ej}\,(10^{-2}M_\odot)$ 
& ~$\langle Y_{e} \rangle$~ & ~$V_{\rm ej}$~ 
& ~$M_{\rm BH}\,(M_\odot)$~ & ~$a_{\rm BH}$~ 
& ~$M_{\rm torus}\,(M_\odot)$~ 
\\
\hline
SFHo-135-135h (high) &(1.35, 1.35) & 1.00 & 150  & 285  & 1.1 & 0.31 & 
0.22 & 2.59 & 0.69 & 0.05 \\
SFHo-135-135l (low)~ &(1.35, 1.35) & 1.00 & 250  & 160  & 1.3 & 0.32 & 
0.21 & 2.60 & 0.70 & 0.03 \\
SFHo-133-137h (high) &(1.37, 1.33) & 0.97 & 150  & 285  & 0.9 & 0.30 & 
0.21 & 2.59 & 0.67 & 0.06 \\
%SFHo133-137l (low)~ &(1.37, 1.33) & 0.97 & 250  & 160  &     & 0.   & 
%& & & \\
SFHo-130-140h (high) &(1.40, 1.30) & 0.93 & 150  & 285  & 0.6 & 0.27 & 
0.20 & 2.58 & 0.67 & 0.09 \\
SFHo-130-140l (low)~ &(1.40, 1.30) & 0.93 & 250  & 160  & 0.6 & 0.27 & 
0.21 & 2.58 & 0.67 & 0.08 \\
SFHo-125-145h (high) &(1.45, 1.25) & 0.86 & 150  & 285  & 1.1 & 0.18 & 
0.24 & 2.58 & 0.66 & 0.12 \\
SFHo-125-145l (low)~ &(1.45, 1.25) & 0.86 & 250  & 160  & 1.2 & 0.19 & 
0.23 & 2.58 & 0.66 & 0.11 \\
DD2-135-135h (high) &(1.35, 1.35) & 1.00 & 160  & 285  & 0.2 & 0.30 &
0.16 &-- & -- & -- \\
DD2-135-135l (low)~ &(1.35, 1.35) & 1.00 & 270  & 160  & 0.2 & 0.30 &
0.15 &-- & -- & -- \\
DD2-130-140h (high) &(1.40, 1.30) & 0.93 & 160  & 285  & 0.3 & 0.26 &
0.18 &-- & -- & -- \\
%DD2 130-140l (low)~ &(1.40, 1.30) & 0.93 & 270  & 160  & 0.  & 0.   &
%&-- & -- & -- \\
DD2-125-145h (high) &(1.45, 1.25) & 0.86 & 160  & 285  & 0.5 & 0.20 &
0.19 &-- & -- & -- \\
%DD2 125-145l (low)~ &(1.45, 1.25) & 0.86 & 270  & 160  & 0.  & 0.   &
%&-- & -- & -- \\
\hline\hline
\end{tabular}
%\end{minipage}
\end{table*}

In numerical simulations, we have to follow the ejecta with the
typical velocity $0.2c$, which expand to
$> 10^3$\,km in $\sim 20$\,ms. To follow the ejecta motion as well as
to resolve neutron stars and merger remnants, we employ a fixed
mesh-refinement algorithm. As in our previous work~\cite{Sekig2015},
we prepare 9 refinement levels with the varying grid spacing as
$\Delta x_l=2^{9-l}\Delta x_9$ ($l=1, 2, \cdots, 9$) and all the
refinement levels have the same coordinate origin. Here, $\Delta x_l$
is the grid spacing for the $l$-th level in Cartesian coordinates.
For each level, the computational domain covers the region $[-N \Delta
  x_{l}, N\Delta x_{l}]$ for $x$- and $y$-directions, and $[0, N\Delta
  x_{l}]$ for $z$-direction (the reflection symmetry with respect to
the orbital plane located at $z=0$ is imposed).  In the
high-resolution run, we assign $N=285$, $\Delta x_9=150$\,m (for the
SFHo EOS) or 160\,m (for the DD2 EOS), and utilize $\approx
7,000$~CPUs on the K computer. Thus the location of outer boundaries
along each axis is $L \agt 10^4$\,km and matter ejected from the central
region never escape from the computational domain in our simulation
time $\lesssim 60$\,ms.  To check whether the numerical results depend
only weakly on the grid resolution, we also performed lower-resolution
simulations for several models.  For this case, $N=160$ and $\Delta
x_9=250$\,m (for the SFHo EOS) or 270\,m (for the DD2 EOS) and we
confirm a reasonable convergence.  We note that since good convergence
is found for the models shown in Table~\ref{tab1}, we do not perform
the low-resolution runs for all the models.  In the following, the
figures are plotted using the results by the high-resolution runs.

%%%%
Choice of the floor density, which has to be put in a dilute-density
or vacuum region outside the neutron stars and merger remnant when
using the conservative form of hydrodynamics in numerical simulations,
is one of the crucial artificial points for accurately exploring the
mass ejection during the merger process. In this study, we set the
floor density to be $1.67 \times 10^4\,{\rm g/cm^3}$. The floor values
of $Y_e$ and temperature are 0.47 and 0.1\,MeV, respectively. For this
case, the artificial floor does not affect the quantitative results of
the mass ejection for $\sim 30$\,ms after the onset of the merger. In
our experiments, we also performed simulations with the floor density
$2\times 10^5\,{\rm g/cm^3}$. In this case, the inertia of the
atmosphere is too high to follow the ejecta motion accurately: The
effect of the atmosphere appeared on the ejecta at $\sim 10$\,ms after
the onset of the merger. In particular for the case that the ejecta
mass is small ($\alt 10^{-3}M_\odot$), this artificial effect is
serious: For example, the ejecta mass steeply decreases with time for
such a low-mass ejecta case because the ejecta are decelerated
significantly.  We find that it is necessary to reduce the floor
density much below $10^5\,{\rm g/cm^3}$ to follow the ejecta for
sufficiently long time until the ejecta motion approximately relaxes
to a free expansion stage~\footnote{Our numerical results for the
  ejecta mass is much larger than those by Ref.~\cite{Palenzuela2015}
  in which simulations are also performed using the SFHo and DD2
  equations of state. We speculate that one of the reasons for this
  would be that our floor density is much smaller than that in
  Ref.~\cite{Palenzuela2015} which employs $5\times 10^5\,{\rm
    g/cm^3}$. See Sec.~III\,B for another reason.}.

%%%%

We consider binary neutron stars with each mass between $1.25M_\odot$
and $1.45M_\odot$ fixing the total mass to be $2.7M_\odot$.  Neutron
stars observed in {\em compact} binary systems typically have the mass
ratio between 0.9 and 1.0, and each mass in the range
$1.23$--$1.45M_\odot$~\cite{Lorimer}. Thus, our choice reasonably
reflects the observational fact. The initial orbital separation is
chosen so that the orbital angular velocity, $\Omega$, satisfies
$Gm_0\Omega/c^3=0.028$ where $m_0$ denotes the total mass, i.e.,
$m_1+m_2=2.7M_\odot$, and $m_1$ and $m_2 (\leq m_1)$ are the mass of
each neutron star in isolation.  Table~\ref{tab1} lists the key
parameters of our models and simulation setup. We define the mass
ratio by $q:=m_2/m_1 (\leq 1)$.

\section{Numerical results}

\subsection{Summary of the merger process}

For all the models we employ in our simulations, a massive neutron
star (MNS) is first formed after the onset of the merger as expected
from our previous results~\cite{Hotoke2013c,Sekig2015} (see also our
earlier papers~\cite{STU05}). The MNS are long-lived in the sense that
their lifetime is much longer than their dynamical time scale and
rotation period $\alt 1$\,ms. The subsequent evolution of the MNS
  depends on the equations of state employed.

For the models with the SFHo EOS, the MNS with mass $\agt 2.6M_\odot$
is hypermassive (see Refs.~\cite{BSS,shibata15} for the
definition of the hypermassive neutron star) because the maximum mass
of spherical and rigidly rotating cold neutron stars is only $\approx
2.06M_\odot$ and $\approx 2.45M_\odot$, respectively, which are smaller
than the remnant MNS mass. As a result, the MNS collapses to a black
hole at $\sim 10$\,ms after the onset of the merger irrespective of
the mass ratio after the angular momentum inside the MNS is
redistributed by the gravitational torque associated with the
non-axial symmetric matter distribution or is dissipated by the
gravitational-wave emission.

The mass and dimensionless spin parameter of the formed black holes
are $\approx 2.6M_\odot$ and $\sim 0.65$--0.70, respectively, and the
remnant black holes are surrounded by a torus with mass $\sim
0.05$--$0.1M_\odot$ and with their typical extent in the orbital plane
$\sim 100$\,km (see Table~\ref{tab1} and Sec.~III\,C for more details).
Such a compact torus would be subsequently evolved by a
magneto-viscous process with the typical lifetime $\tau_v \sim
(\alpha_v \Omega)^{-1}$ where $\alpha_v$ is the so-called
$\alpha$-parameter for viscous hydrodynamics and $\tau_v \sim
10^2\,{\rm ms}\,(\alpha_v/10^{-2})^{-1}$ for $\Omega=O(10^3\,{\rm
  rad/s})$~(see, e.g., Ref.~\cite{kiuchi2014}). Thus, for a plausible
value of $\alpha_v \sim 0.01$, this system is a candidate for the
central engine of short-hard gamma-ray bursts with the duration less
than one second, like GRB\,130603B~\cite{GRB130603B} (see also
Sec.~III\,E).

For the DD2 case, any of the formed MNS does not collapse to a black
hole in our simulation time $\sim 50$\,ms. This is reasonable because
the maximum mass of spherical and rigidly rotating cold neutron stars
for the DD2 EOS is high, i.e., $\approx 2.42M_\odot$ and $2.8M_\odot$,
respectively, and hence, the formed hot MNS with mass $\sim
2.6M_\odot$ are not hypermassive and cannot collapse to a black hole
before a substantial fraction of the angular momentum and thermal
energy are dissipated or carried away, respectively, by some
angular-momentum transport processes and the neutrino emission (for
which the cooling time scale is longer than 1\,s; e.g.,
Refs.~\cite{Sekig,Hotoke2013c}). The hot remnant MNS would be
long-lived with their lifetime longer than a few seconds and could be
a strong emitter of neutrinos, which can modify the chemical property
of the ejecta via the neutrino irradiation process (see Sec.~III\,C).

\subsection{Dynamical mass ejection}

%% FIG1: q=1 & 0.86

\begin{figure*}[t]
%\epsfxsize=3.2in
%\leavevmode
%\epsffile{Mej-aveye_SFHo.eps}
\includegraphics[scale=0.76]{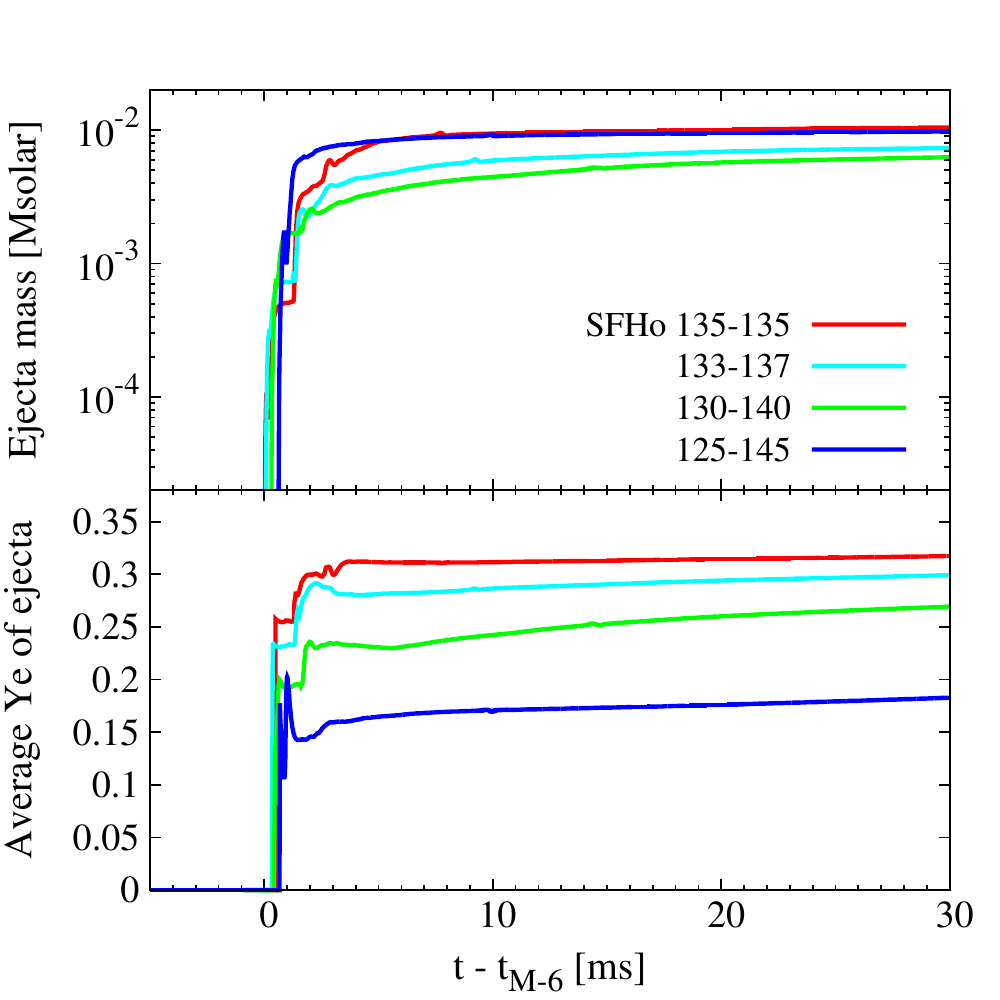}
\hspace{10mm}
%\epsfxsize=3.2in
%\leavevmode
%\epsffile{Mej-aveye_DD2.eps}
%\epsffile{fig1b.eps}
\includegraphics[scale=0.76]{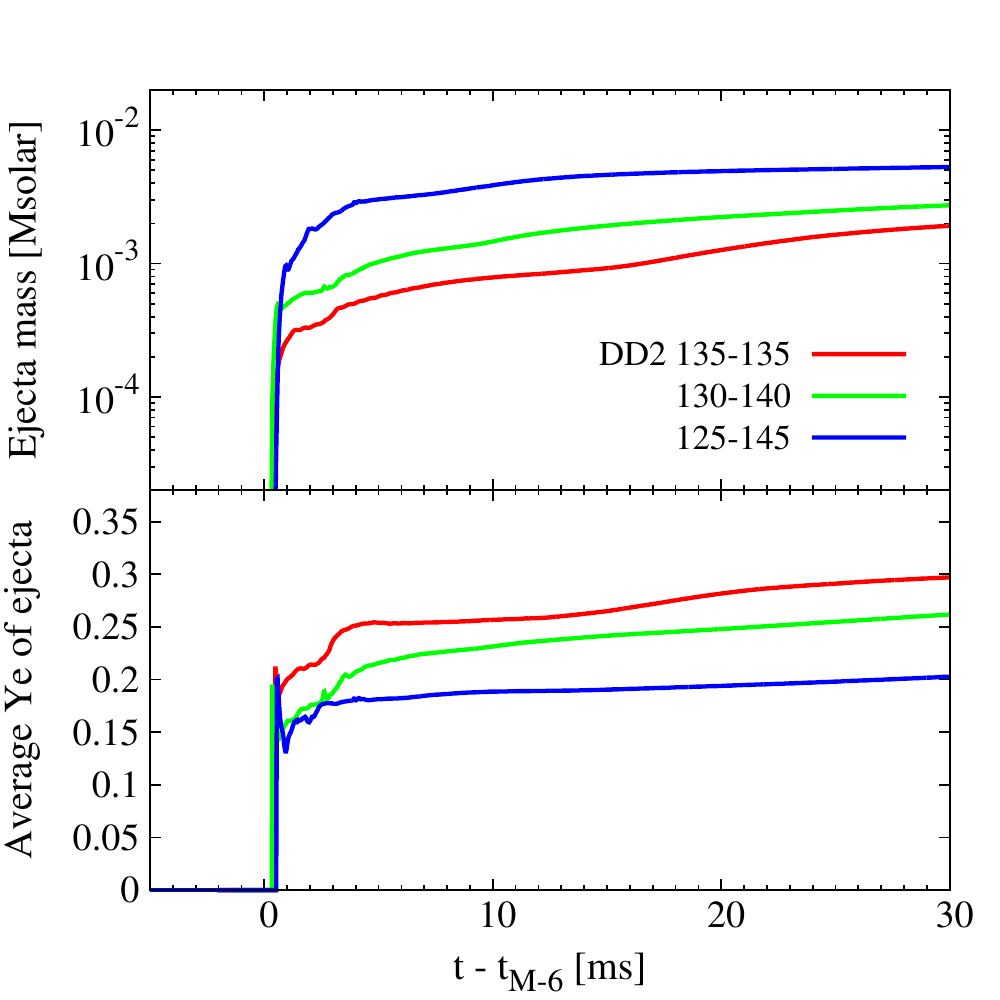}
\caption{Rest mass (upper panel) and averaged value of $Y_e$ (lower
  panel) of the dynamical ejecta as functions of time for the SFHo
  (left) and DD2 (right) models and for a variety of binary mass
  ratios. $t_{M_{-6}}$ approximately denotes the time at the onset of
  the merger (see text). The results for the high-resolution runs are
  plotted.  A substantial fraction of the matter is dynamically
  ejected at $t - t_{M_{-6}} \alt 2$\,ms but gradual ejection continues 
  subsequently. The long-term gradual increase of the ejecta mass and
  the averaged value of $Y_e$ for $t_{M_{-6}} \agt 10$\,ms, observed
  in particular for the DD2 case, is due to the irradiation by
  neutrons that are emitted from the merger remnant.
\label{fig1}}
\end{figure*}

%% FIG2 something that clarifies the tidal & shock heating ejection

\begin{figure*}[t]
\begin{center}
    \includegraphics[scale=0.5]{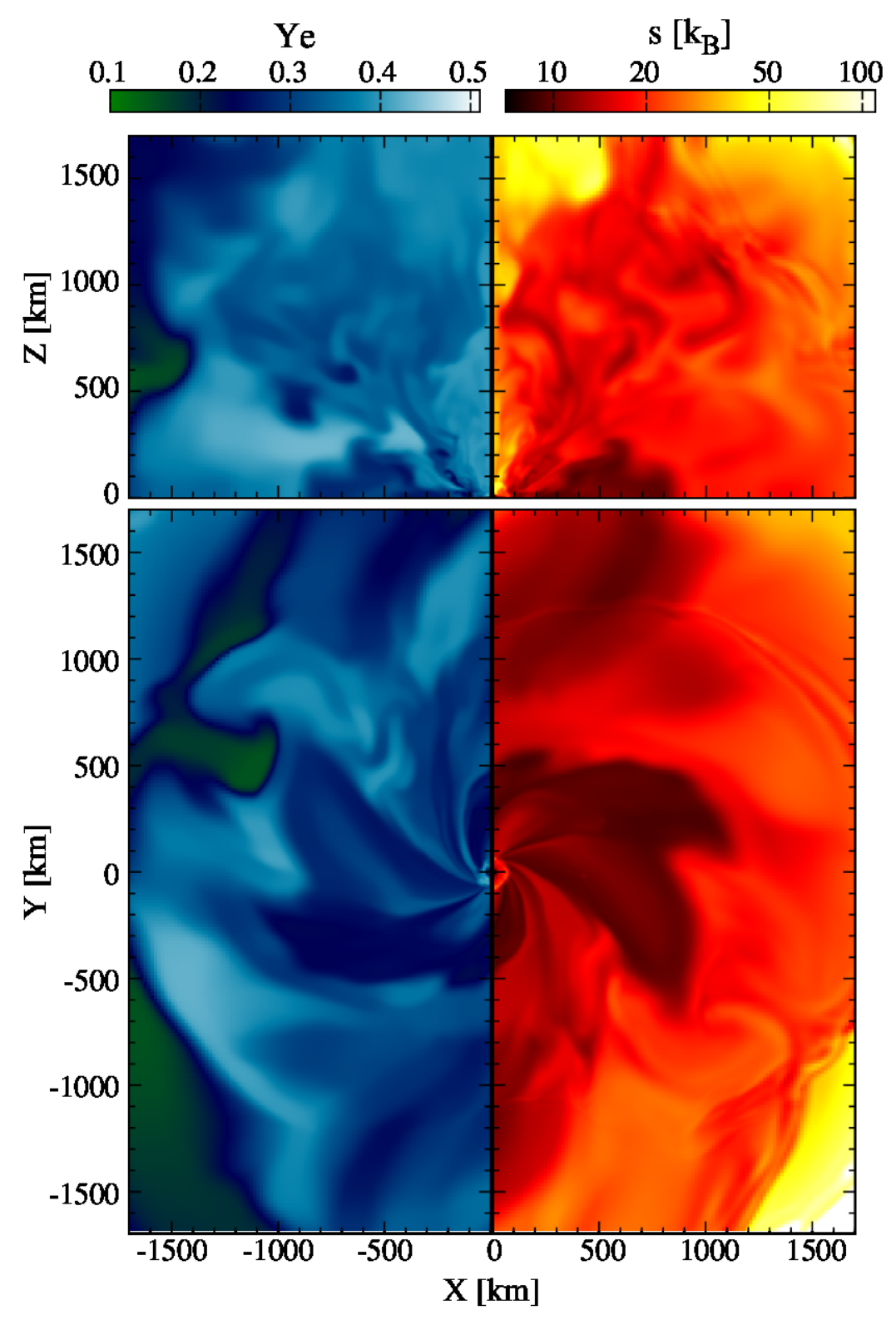}
    \includegraphics[scale=0.5]{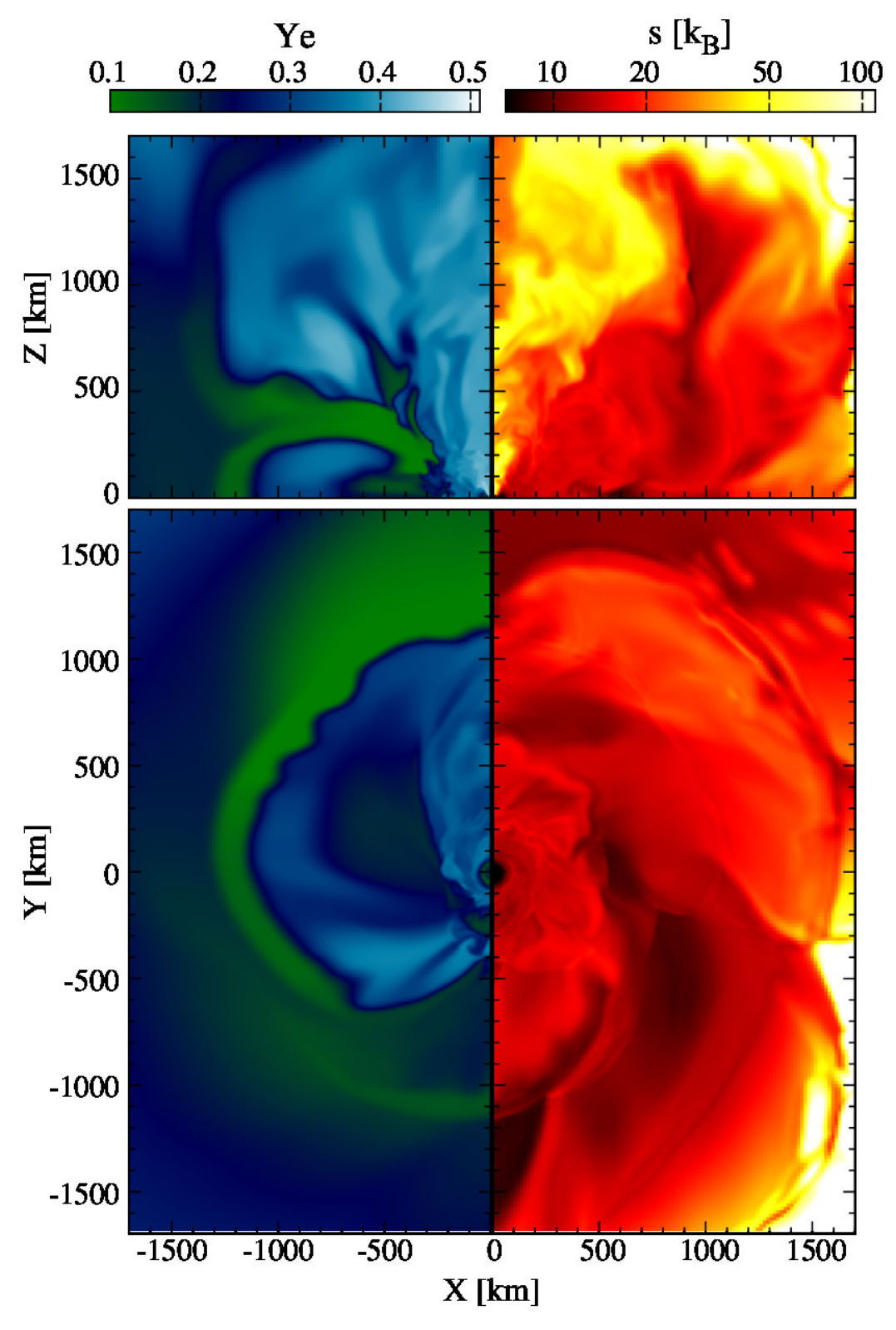}
    \includegraphics[scale=0.5]{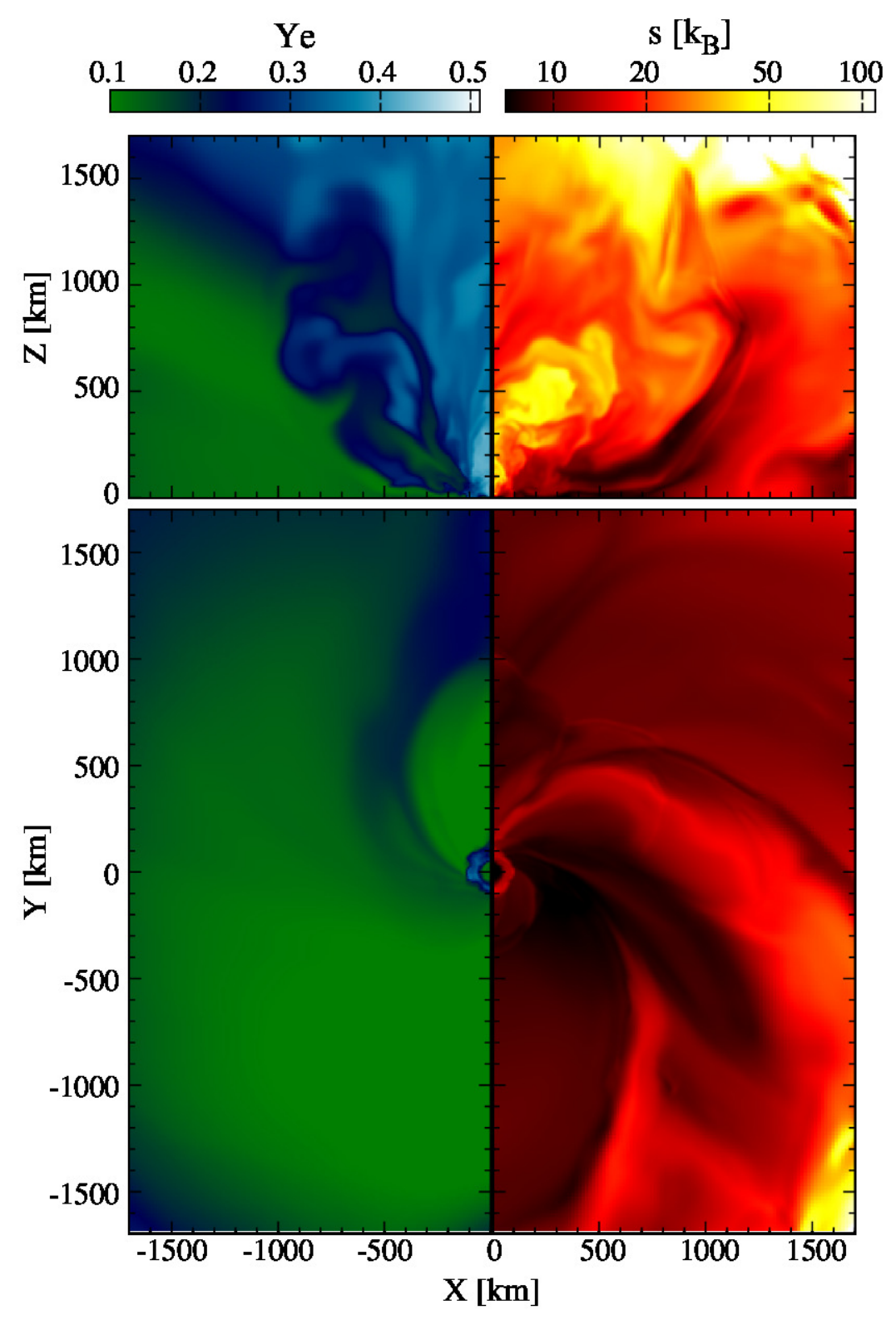} \\
    \includegraphics[scale=0.5]{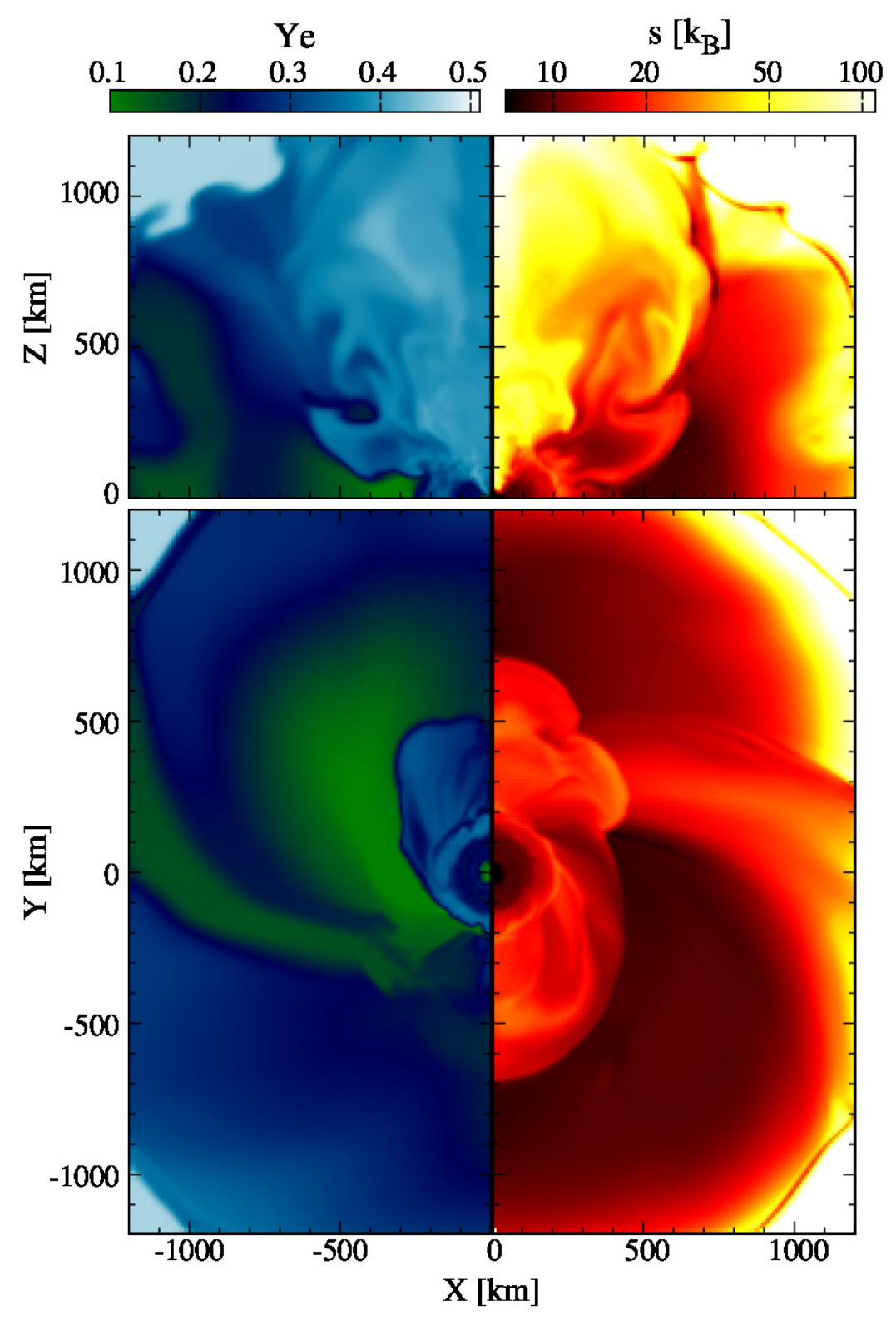}
    \includegraphics[scale=0.5]{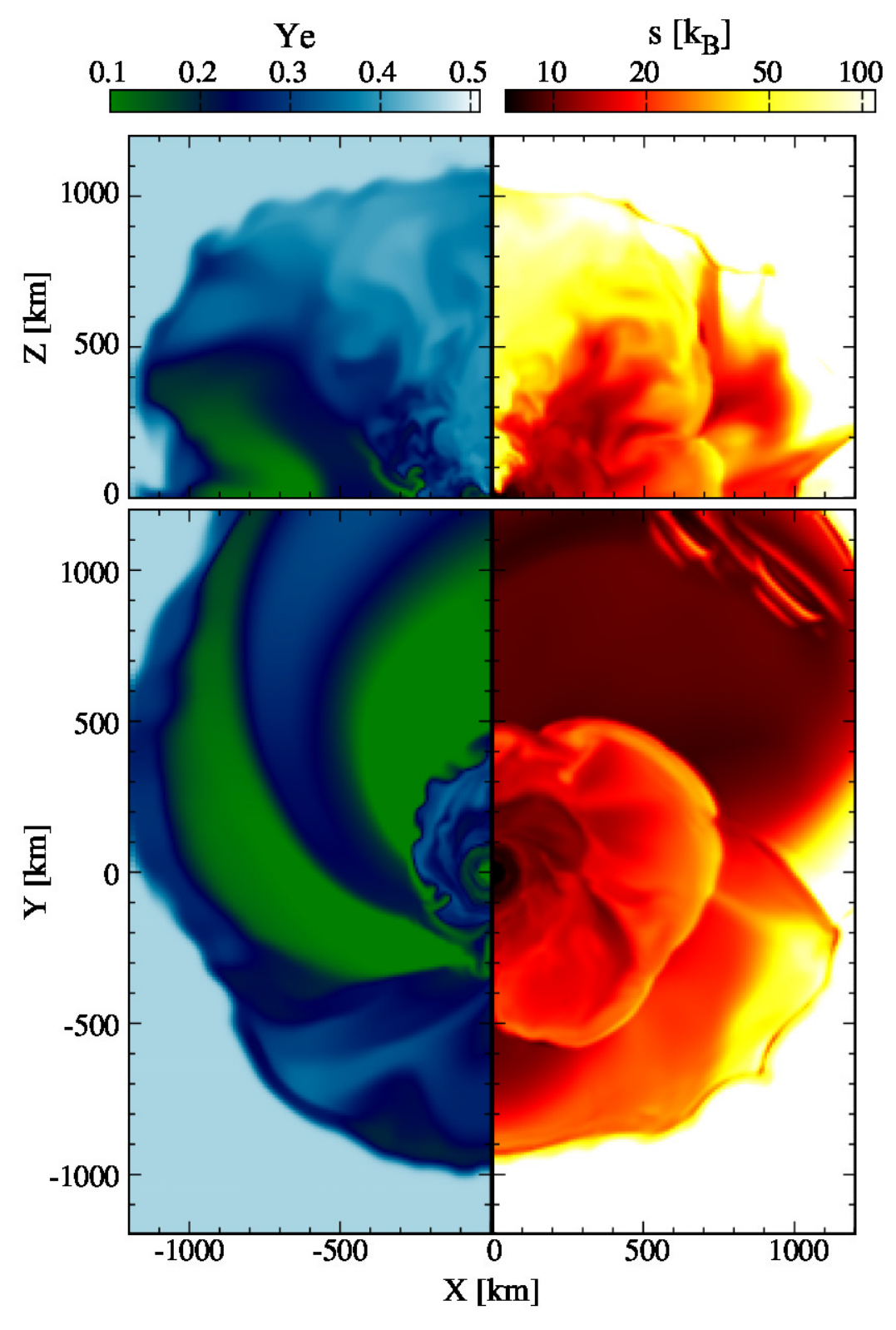}
    \includegraphics[scale=0.5]{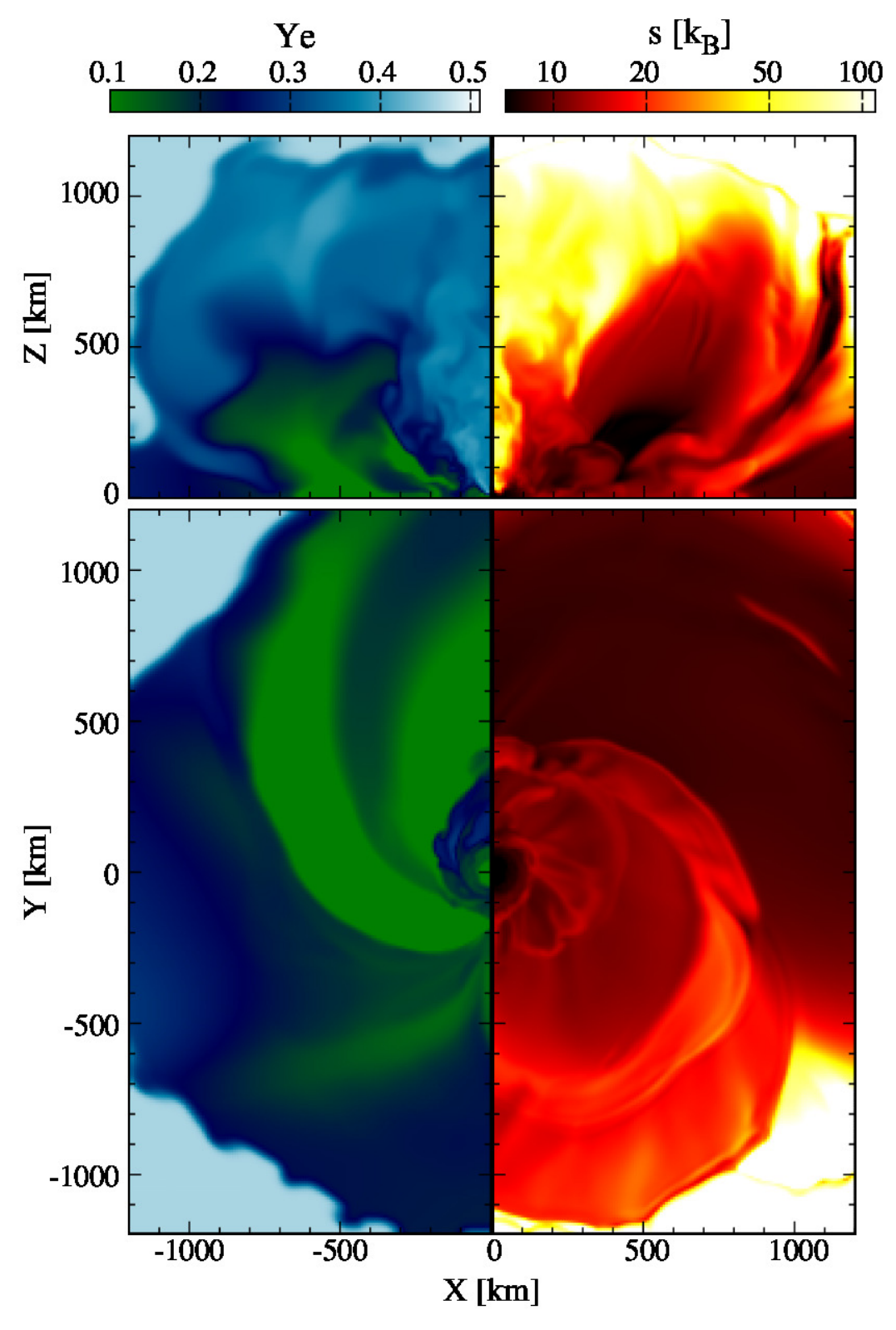}
\end{center}
\vspace{-3mm}
\caption{Profiles of the electron number per baryon, $Y_{e}$, (left in
  each panel) and the specific entropy, $s$, (right in each panel) in
  $x$-$y$ (lower in each panel) and $x$-$z$ (upper in each panel)
  planes. The top three panels show the results for SFHo-135-135h
  (left), SFHo-130-140h (middle), and SFHo-125-145h (right) at
  $\approx 13$\,ms after the onset of the merger.  The lower three
  panels show the results for DD2-135-135h (left), DD2-130-140h
  (middle), and DD2-125-145h (right) at $\approx 10$\,ms after the
  onset of the merger.
\label{fig2}}
\end{figure*}

\begin{figure*}[t]
\includegraphics[scale=0.76]{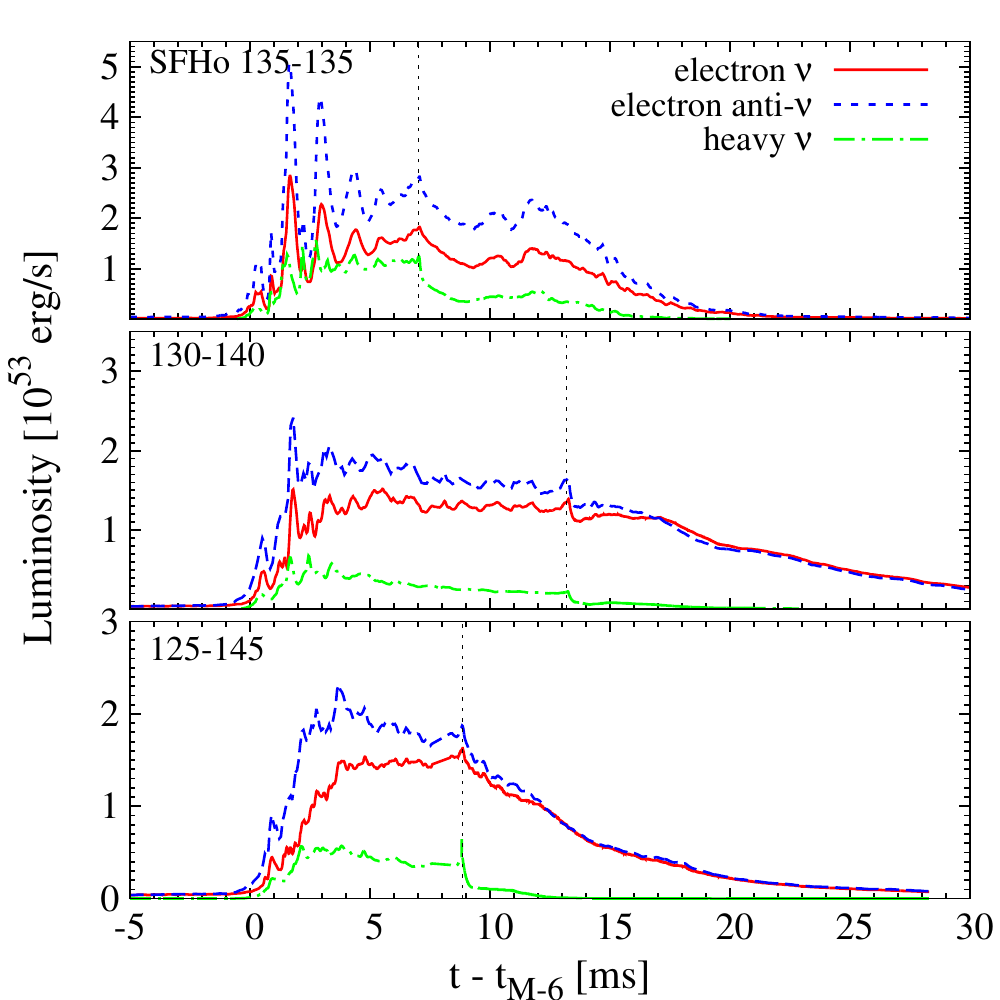}
\hspace{8mm}
\includegraphics[scale=0.76]{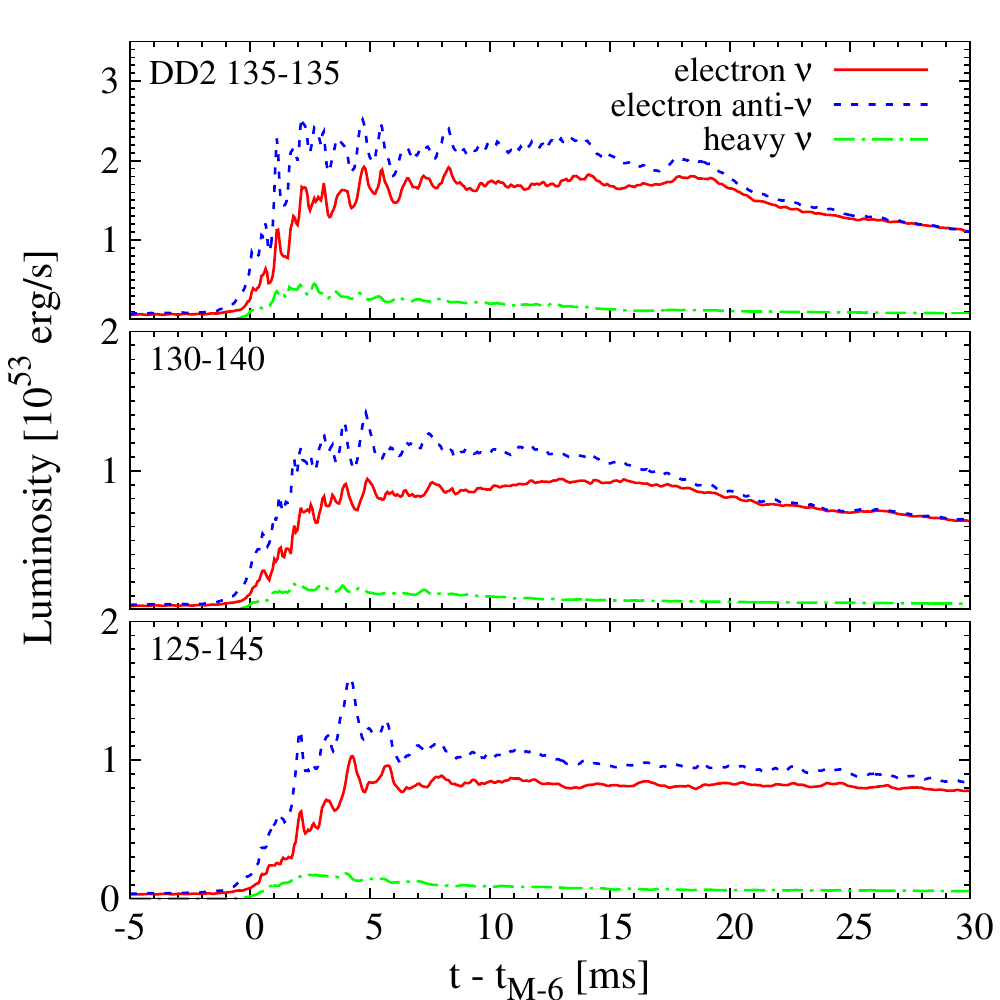}
\caption{Luminosity curves of $\nu_{e}$ (red solid), $\bar{\nu}_{e}$
  (blue dashed), and heavy (green dotted-dashed) neutrinos for the
  models with the SFHo EOS (left) and the DD2 EOS (right),
  respectively (note that the scales in the vertical axis are
  different among the plots).  For heavy neutrinos, the contribution
  from only one heavy species is plotted.  The vertical dashed lines
  in the left panel shows the time at the formation of a remnant black
  hole.  We note that the relatively high heavy-neutrino luminosity
  for the SFHo models before the collapse to the remnant black holes
  reflects the fact that the temperature of remnant MNS is higher and
  the pair-process neutrino emission is more active than those for the
  DD2 model.
\label{fig3}}
\end{figure*}

%%%%%%%%%%%%%%%%%%% We have to choose mass ratio
%%%%%%%%%%% q=1 & 0.86 

Figure~\ref{fig1} plots the evolution of the total rest mass, $M_{\rm
  ej}$, and the averaged value for the electron number per baryon,
$\langle Y_e \rangle$, of the ejecta for the models with the SFHo and
DD2 EOS for a variety of mass ratios.  Here, $t_{M_{-6}}$
approximately denotes the time at the onset of the merger: It denotes
the time at which $M_{\rm ej}$ exceeds $10^{-6}M_{\odot}$. The average
of $Y_e$ for the ejecta is defined by
\begin{equation}
\langle Y_e \rangle ={1 \over M_{\rm ej}} \int Y_e \,dM_{\rm ej}. 
\end{equation}
We specify the matter as the ejecta if the lower time component of the
fluid four velocity, $u_t$, is smaller than $-1$ as
before~\cite{Sekig2015}. We note that this condition agrees
approximately with the condition $h u_t < -1$ where $h$ is the
specific enthalpy. The reason for this is that $h$ is close to unity
for the ejecta components moving far from the merger remnant located
in the central region. In Table~\ref{tab1}, we also summarize the
total rest mass, the averaged value of $Y_e$, and the averaged
velocity of the ejecta, $V_{\rm ej}$, all of which are measured at
$t-t_{M_{-6}}\approx 30$\,ms.  Here, $V_{\rm ej}$ is defined by
$\sqrt{2E_{\rm kin}/M_{\rm ej}}$ where $E_{\rm kin}$ is total kinetic
energy of the ejecta.

Figure~\ref{fig1} illustrates that the ejecta mass depends strongly on
the EOS employed, as already described in Ref.~\cite{Sekig2015}: For
the smaller value of $R_{1.35}$, the ejecta mass is larger (see also
Ref.~\cite{Palenzuela2015}).  Figure~\ref{fig1} also shows that for
the models with the SFHo EOS, the ejecta mass depends weakly on the
binary mass asymmetry, while for those with the DD2 EOS, it increases
steeply with the increase of the degree of the binary mass
asymmetry. As already described in our study of
Ref.~\cite{Hotoke2013b} in which piecewise polytropic EOS is
  employed, this is due to the fact that there are two major
dynamical mass ejection mechanisms (see also
Ref.~\cite{Bauswein}): shock heating and tidal interaction (i.e.,
tidal torque exerted by elongated two neutron stars and highly
non-axisymmetric merger remnants).  For the equal-mass or slightly
asymmetric case, the shock heating is the primary player of the
dynamical mass ejection for neutron stars with soft EOS like the SFHo
EOS, while the tidal torque is the primary player for binary neutron
stars with stiff EOS like the DD2 EOS.

The shock heating efficiency during the merger phase decreases with
the increase of the binary asymmetry degree because the smaller-mass
neutron star in such asymmetric systems is tidally elongated just
prior to the onset of the merger, avoiding the coherent collision with
the heavier companion at the merger.  Thus, for the models with the
SFHo EOS, the shock heating effect is weakened with the increase of
  the binary asymmetry degree while the importance of the tidal effect
  is enhanced.  As a result of this change in the dynamical mass
ejection mechanism, the ejecta mass slightly decreases with the change
of $q$ from unity to a smaller value to $\sim 0.9$.  However, with
the further decrease of $q$ (i.e., with the further increase of the
degree of the mass asymmetry), the ejecta mass increases because the
enhanced tidal effect dominates the reduced shock heating effect.

On the other hand, for the DD2 models the tidal interaction is always
the primary mechanism for the dynamical mass ejection. The importance
of the tidal effect is further enhanced with the increase of the mass
asymmetry degree for this EOS, monotonically increasing the dynamical
ejecta mass. Thus, for significantly asymmetric binaries, the typical
ejecta mass would approach $10^{-2}M_\odot$ irrespective of the EOS
employed.  We note that the total ejecta mass depends only weakly on
the grid resolution as listed in Table~\ref{tab1}.

%%%

As shown in Fig.~\ref{fig1}, the ejecta mass increases with time for
the first $\sim 10$\,ms after the onset of the merger. This is in
particular observed for the SFHo models with $q \agt 0.9$ and all the
DD2 models.  This indicates that we have to follow the ejecta motion
at least for $\approx 10$\,ms after the onset of the merger.  In a
recent simulation of Ref.~\cite{Palenzuela2015}, they estimated the
properties of the ejecta at $\alt 5$\,ms after the onset of the
merger, perhaps because of their small computational domain employed
($L=750$\,km).  However, the ejecta mass would still increase with
time in such an early phase. This could be one of the reasons that our
results for the ejecta mass are much larger than theirs.
Figure~\ref{fig1} also shows that the average of $Y_e$ still
significantly varies with time for the first $\sim 5$\,ms after the
onset of the merger. This also shows that it would be necessary to
determine the properties of the ejecta at $\agt 10$\,ms after the
onset of the merger (if the average of $Y_e$ is estimated at $\sim
5$\,ms after the onset of the merger as in Ref.~\cite{Palenzuela2015},
the average of $Y_e$ could be underestimated).

Irrespective of the EOS and mass ratios, the averaged ejecta velocity
is in the range between $0.15c$ and $0.25c$, as found in
Refs.~\cite{Hotoke2013b,Sekig2015,radice2016}. As we already pointed
out in Ref.~\cite{Hotoke2013b}, the ejecta velocity is higher for
softer EOS and this shows that the shock heating effect enhances the
ejecta velocity.  On the other hand, the ejecta velocity depends only
weakly on the mass ratio (as long as it is in the range $0.85 < q \leq
1$), although it is slightly increased for significantly asymmetric
binaries like 1.25--$1.45M_\odot$ models.

%%%%%%%%% FIG2

As described earlier in this section, shock heating and tidal
interaction are two major dynamical mass ejection mechanisms.  By the
tidal torque, the matter tends to be ejected near the orbital plane
because the tidal-force vector primarily points to the direction in
this plane.  On the other hand, by the shock heating, the matter is
ejected in a quasi-spherical manner like in supernova explosion.
Because both effects play a role, the dynamical ejecta usually have a
spheroidal morphology~\cite{Hotoke2013b}.  

For the SFHo models, the shock heating plays a primary role for the
equal-mass or slightly asymmetric case, and hence, the dynamical
ejecta in this case have a quasi-spherical morphology. However, for
the significantly asymmetric case, e.g., with $q \sim 0.85$, the tidal
effect becomes appreciable, as already mentioned, and hence, the
anisotropy of the dynamical ejecta is enhanced.  On the other hand,
for the DD2 models, the tidal torque always plays a primary role for
the dynamical mass ejection. Thus, with the increase of the binary
asymmetry degree, this property is further enhanced, and the
anisotropy of the dynamical ejecta morphology is increased. Here, we
note that the degree of the anisotropy is correlated with the
neutron-richness of the dynamical ejecta because (i) the tidally
ejected components are less subject to the thermal
weak-interaction reprocess associated with the shock heating
preserving the neutron-rich nature of the original neutron-star
  matter and (ii) the neutrino irradiation is less subject to the
matter near the equatorial plane than that near the polar region
(see the discussion in Sec.~III\,C).

%%%%%%% 

Six panels of Fig.~\ref{fig2} display the profiles of the electron
number per baryon, $Y_e$, (left side of each panel) and specific
entropy, $s$, (right side of each panel) of the ejecta on the $x$-$y$
and $x$-$z$ planes for the SFHo (top panels) and DD2 (lower panels)
models. For the SFHo and DD2 models, the snapshots at
$t-t_{M_{-6}}\approx 13$\,ms and 10\,ms are plotted, respectively.
The left, middle, and right panels display the results for
1.35-$1.35M_\odot$, 1.30-$1.40M_\odot$, and 1.25-$1.45M_\odot$,
respectively.  This figure shows a clear dependence of the properties
of the dynamical ejecta on the binary asymmetry degree and on the EOS
employed as follows:

\noindent
(I) For the SFHo models, the specific entropy of the ejecta decreases
steeply with the increase of the binary asymmetry degree in
  particular near the orbital plane.  This is due to the fact that
the effect of the shock heating at the onset of the merger, which
contributes a lot to the dynamical mass ejection, becomes weak
with the increase of the binary asymmetry degree.

\noindent
(II) As a result, for the SFHo models, the ejecta component with low
values of $Y_e$ increases with the increase of the binary asymmetry
degree: For the equal-mass case, the ejecta with $Y_e \gtrsim 0.2$ are
the primary components while for the $1.25$--$1.45M_\odot$ model,
those with $Y_e \lesssim 0.2$ are primary (in particular for the
components near the orbital plane).  This is due to the following
fact: For a high temperature environment, $e^-e^+$ pair-creation is
enhanced, and consequently, the positron capture reaction,
  $n+e^+\rightarrow p+\bar\nu_e$, efficiently proceeds in neutron-rich
  matter, resulting in the increase of $Y_e$.  With the increase of
the binary asymmetry degree, the shock heating effect becomes less
important and the temperature for a substantial fraction of the
dynamical ejecta is decreased. As a result, the positron production
and resulting positron capture are suppressed. Hence, the
neutron richness is preserved to be relatively high (the value of
$Y_e$ is preserved to be low).

\noindent
(III) For the DD2 models, the effect  associated with the binary asymmetry
found for the SFHo model is not very remarkable: The typical values of
$Y_e$ and specific entropy depend mildly on the binary asymmetry
degree, although we still observe a monotonic decrease of these values
(see, e.g., Fig.~1).  This weak dependence is due to the fact that the
ejecta are composed primarily of tidally-ejected matter irrespective
of the mass ratio, as already mentioned. 

\subsection{Neutrino irradiation}

For the DD2 models, the remnant massive neutron stars are long-lived,
while for the SFHo models, the remnants collapse to a black hole in
$\sim 10$\,ms after the onset of the merger. Therefore, a
high-luminosity neutrino emission is continued for a long time scale
from the remnant of the DD2 models, while the strong emission
continues only briefly for the SFHo models (see Fig.~3).  As a result,
a long-term {\em neutrino-irradiation}
effect~\cite{Dessart,Perego2014,Just2015,martin,Sekig2015} plays an
important role for heating up the ejecta and for increasing the value
of $Y_{e}$ (see Fig.~1), in particular in the region above the remnant
MNS pole (see Fig.~2) in the DD2 model.

As we pointed out in Ref.~\cite{Sekig2015}, the reason for the
increase of $Y_e$ by the neutrino irradiation is as follows: The
luminosity of electron neutrinos emitted from the remnant hot MNS is
quite high as shown in Fig.~3.  In such an environment, neutrino
capture processes, $n+\nu_e \rightarrow p + e^-$ and $p + \bar\nu_e
\rightarrow n + e^+$, are activated in the matter surrounding the MNS.
By the balance of these reactions, the fractions of neutrons and
protons are determined and the equilibrium value of $Y_{e}$ will
  be given by (e.g., Ref.~\cite{Qian1996}),
\begin{equation}
Y_{e, \rm{eq}} \sim \left[ 1 + \frac{L_{\bar{\nu}_{e}}}{L_{\nu_{e}}}\cdot
\frac{\langle \epsilon_{\bar{\nu}_{e}} \rangle - 2\Delta}
{\langle \epsilon_{\nu_{e}} \rangle + 2\Delta} \right]^{-1},
\end{equation}
where $\Delta = m_{n}c^{2}-m_{p}c^{2} \approx 1.293$ MeV, $\langle
\epsilon_{\nu_{e}} \rangle$ and $\langle \epsilon_{\bar{\nu}_{e}}
\rangle$ denote averaged neutrino energy of $\nu_e$ and $\bar\nu_e$,
and $L_{\nu_e}$ and $L_{\bar\nu_e}$ denote the luminosity of $\nu_e$
and $\bar\nu_e$, respectively. For the DD2 models, $\langle
\epsilon_{\nu_{e}}\rangle \approx 10$ MeV, $\langle
\epsilon_{\bar{\nu}_{e}}\rangle \approx 15$ MeV, and
$L_{\bar{\nu}_{e}}/L_{\nu_{e}}\approx 1.0$--1.3, and consequently, the
equilibrium value is $Y_{e} \approx 0.45$--0.5.  Due to the neutrino
irradiation, the neutron richness of the originally neutron-rich
matter with $Y_e \lesssim 0.1$ is decreased (the average of $Y_e$ is
increased) towards the equilibrium value.

However, this neutrino irradiation effect depends on the binary
asymmetry because, as Fig.~\ref{fig3} shows, the neutrino luminosity
decreases with the increase of the binary asymmetry degree (this is in
particular seen clearly among the DD2 models).  A time scale for
  the increase of the average $Y_{e}$ may be estimated approximately 
  as
\begin{eqnarray}
\tau_{Y_{e}} &\sim&
\langle Y_{e}\rangle \left[\frac{1}{4\pi r^{2}}
\left( \frac{X_{n}\sigma_{\nu_{e}n}L_{\nu_{e}}}{\langle \epsilon_{\nu_{e}} \rangle} -
\frac{X_{p}\sigma_{\bar{\nu}_{e}p}L_{\bar{\nu}_{e}}}
{\langle \epsilon_{\bar{\nu}_{e}} \rangle}\right)
 \right]^{-1} \nonumber \\
&\sim& 40\,{\rm ms}\,
\bigg(\frac{L_{\nu}}{10^{53}\,{\rm ergs/s}}\bigg)^{-1}
\bigg(\frac{r}{100\,{\rm km}}\bigg)^{2},
\end{eqnarray}
where $r$ is the coordinate radius, $\sigma_{\nu_{e}n}$ and
$\sigma_{\bar{\nu_{e}}p}$ are the cross-sections of the $\nu_{e}$
absorption on neutrons and $\bar{\nu}_{e}$ on protons, respectively.
Here, we set $\langle \epsilon_{\nu_{e}}\rangle = 10$ MeV, $\langle
\epsilon_{\bar{\nu}_{e}}\rangle = 15$ MeV, $L_{\nu_{e}} =
L_{\bar{\nu}_{e}} = L_\nu$, $X_{n}=1-\langle Y_{e} \rangle$, and
$X_{p}=\langle Y_{e}\rangle$ with $\langle Y_{e}\rangle = 0.2$. 
Thus, for the asymmetric binaries for which $L_\nu$ is smaller than
that for the equal-mass binary, the time scale to increase $Y_e$ by
the neutrino irradiation is longer, as found in Fig.~1: It shows that
the rate for the long-term increase in $\langle Y_e \rangle$ is
smaller for the more asymmetric binary models.

%%%%%%%%%%% FIG3: Neutrino luminosity & dependence on q

By this neutrino irradiation, the ejecta mass is also increased (see
Fig.~1). This is in particular the case for the DD2 models with the
equal-mass or weakly asymmetric systems, for which the remnant MNS is
long-lived and a long-term increase of the ejecta component is
observed. For the SFHo models, the MNS is hypermassive and collapses
to a black hole in $\sim 10$\,ms after the onset of the merger,
reducing the neutrino luminosity. Thus, the effect of the neutrino
irradiation is less important irrespective of the binary asymmetry
degree.

\begin{figure}[t]
\includegraphics[scale=0.80]{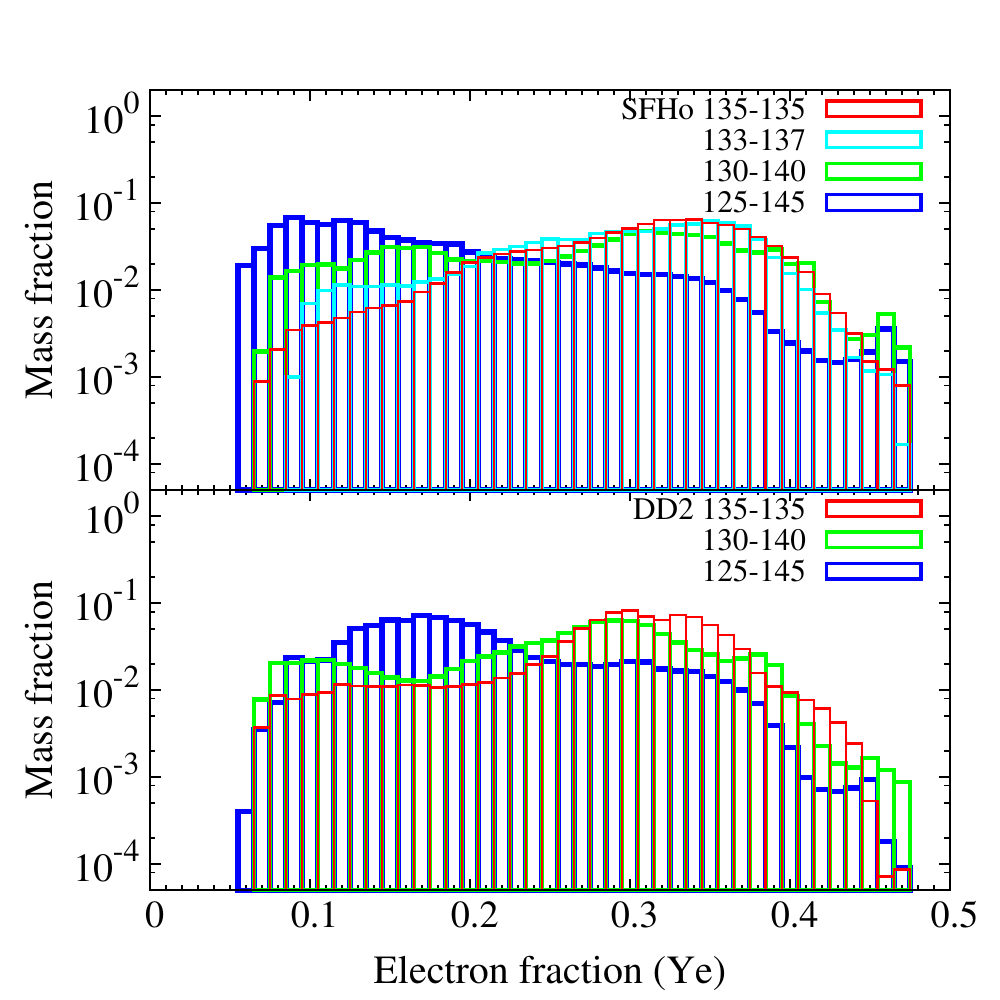}
\caption{The mass-distribution histograms with respect to $Y_e$
  normalized by the total mass of ejecta for the models with the SFHo
  EOS (top panel) and the DD2 EOS (lower panel), respectively.  The
  data at $t-t_{M_{-6}}\approx 25$\,ms are employed.
\label{fig4}}
\end{figure}

%%%%%%%% histogram%%%%% Y_e vs q SFHo

\subsection{Mass distribution of $Y_{e}$}

The effect of the binary asymmetry is also reflected in the mass
distribution of $Y_e$ in an appreciable manner in particular for the
SFHo models. Figure~\ref{fig4} shows histograms for the ejecta mass
fraction as a function of $Y_e$ at $t-t_{M_{-6}}\approx 25$\,ms, at which
the total (dynamical) ejecta mass and the averaged value of $Y_e$
approximately settle to relaxed values.  

For the equal-mass or slightly asymmetric cases with the SFHo EOS, the
ejecta typically have high values of the specific entropy due to
strong shock heating at the onset of the merger (see Fig.~2). As a
result of this high value (i.e., the high value of temperature),
$e^-e^+$ pair-creation is enhanced and subsequently positron capture,
$n+e^+ \rightarrow p+\bar\nu_e$, efficiently proceeds, resulting in
the increase of $\langle Y_e \rangle$.  Because the shock heating
effect for the SFHo models is more significant than that for the DD2
models, the averaged value of $Y_e$ for the ejecta of the SFHo models
is higher than that of the DD2 models for the equal-mass or slightly
asymmetric cases (see Fig.~\ref{fig1}).

On the other hand, in the presence of appreciable binary asymmetry,
not only the shock heating but also the tidal effect become important
in the dynamical mass ejection even for the SFHo models. As a result,
the fraction of matter with low values of $Y_e$ is increased. This is
clearly observed in Fig.~\ref{fig4}, which shows that the value of
$Y_e$ at the peak gradually shifts to the lower-value side and in
particular for the $1.25$--$1.45M_\odot$ model, the peak $Y_e$ value
is smaller than 0.2 both for the SFHo and DD2 models.  However, even
in such appreciably asymmetric cases, the dynamical ejecta have a
broad distribution in $Y_e$. This is the universal qualitative feature
and well-suited for producing a variety of $r$-process heavy
elements~\cite{Wanajo}.

\subsection{Properties of the merger remnant}

\begin{figure*}[pt]
\begin{center}
    \includegraphics[scale=0.64]{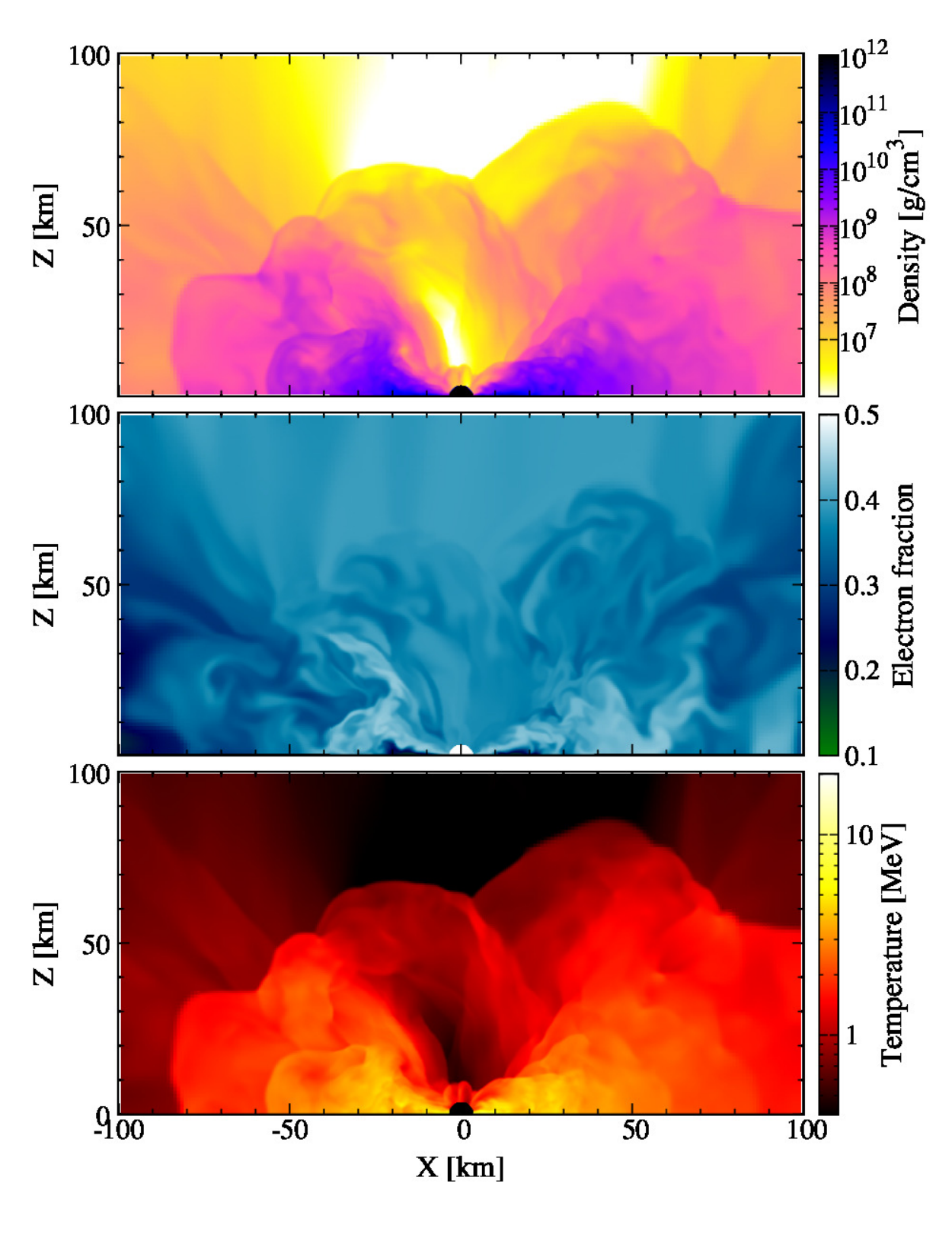}
\hspace{8mm}
    \includegraphics[scale=0.64]{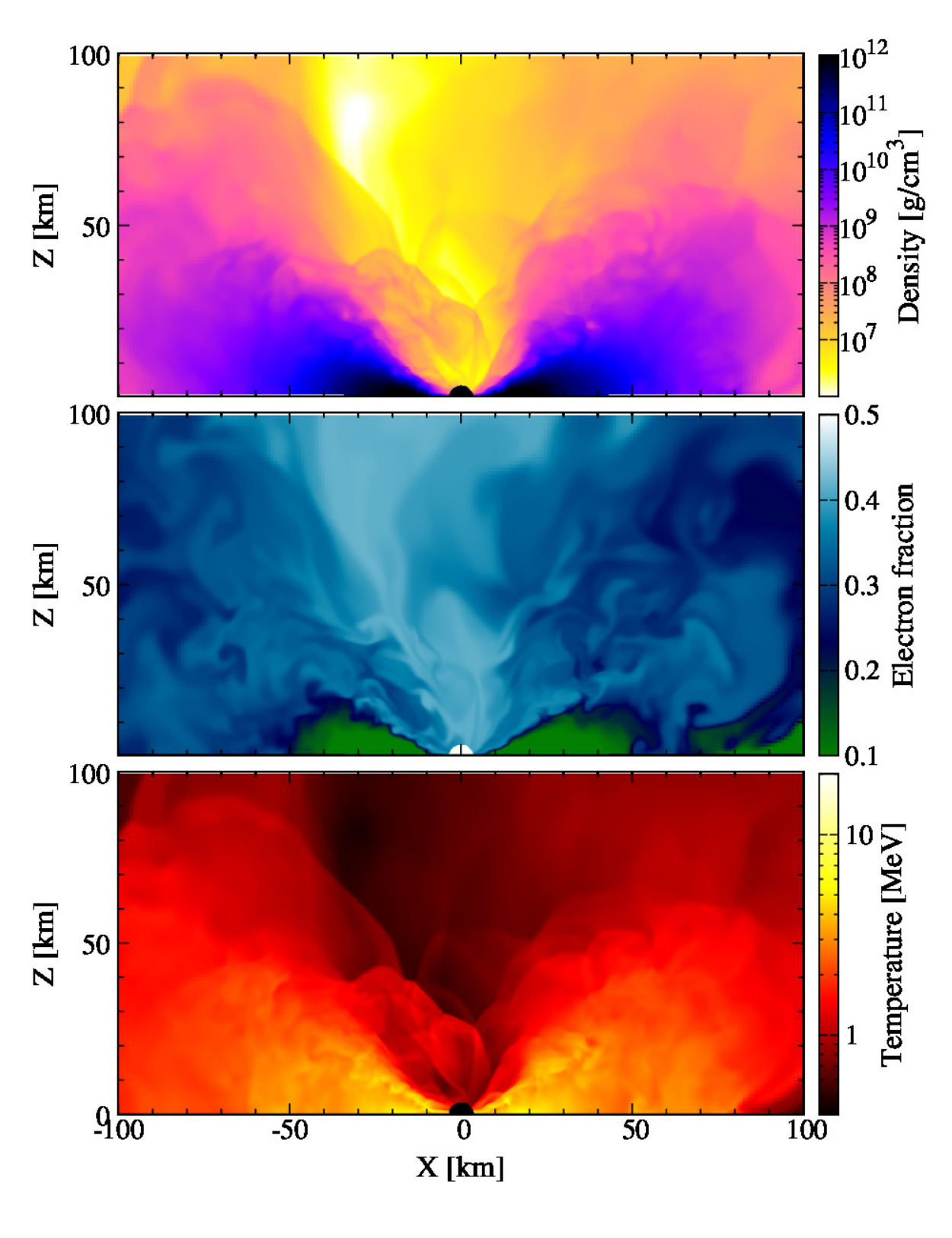} \\
    \includegraphics[scale=0.32]{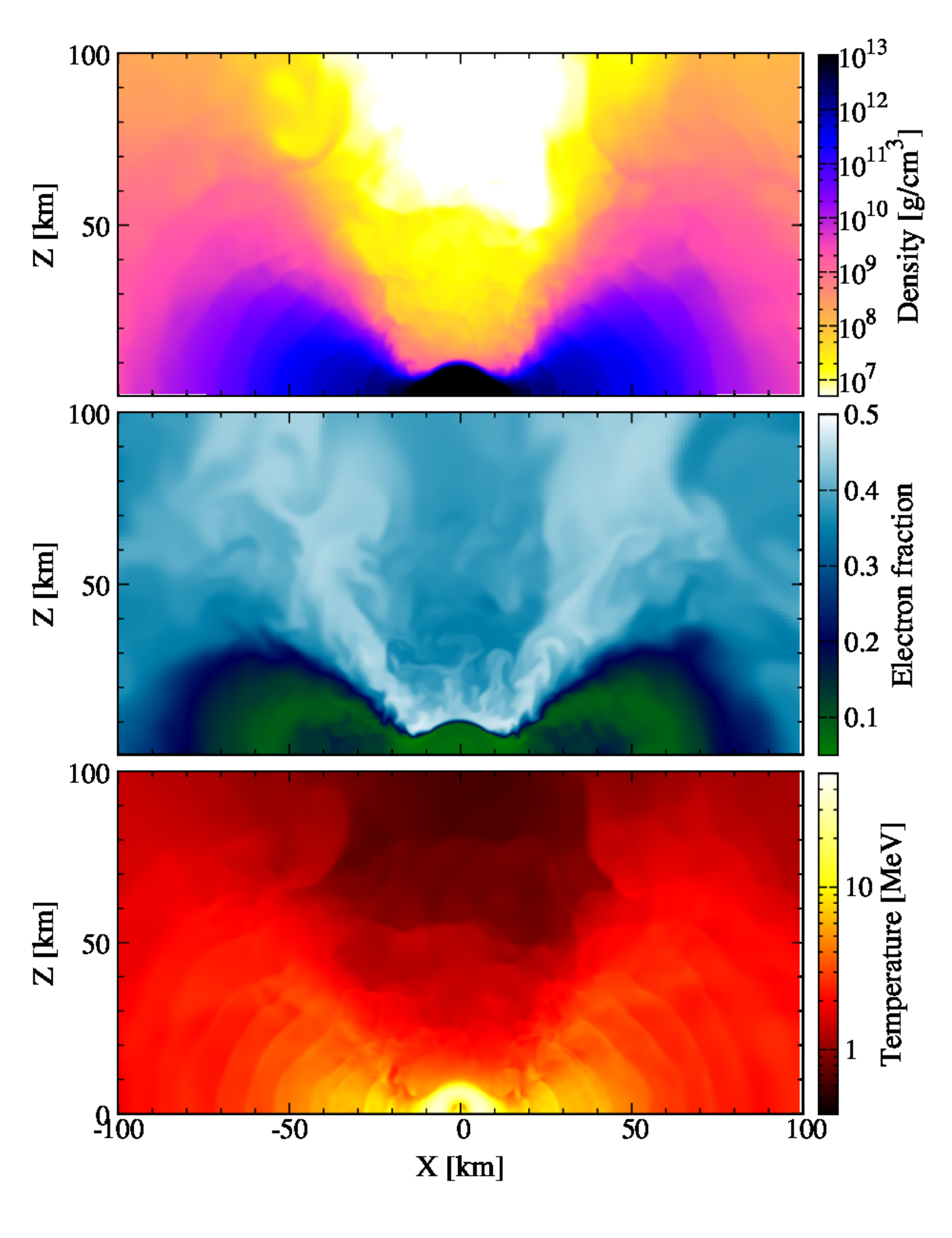}
\hspace{8mm}
    \includegraphics[scale=0.64]{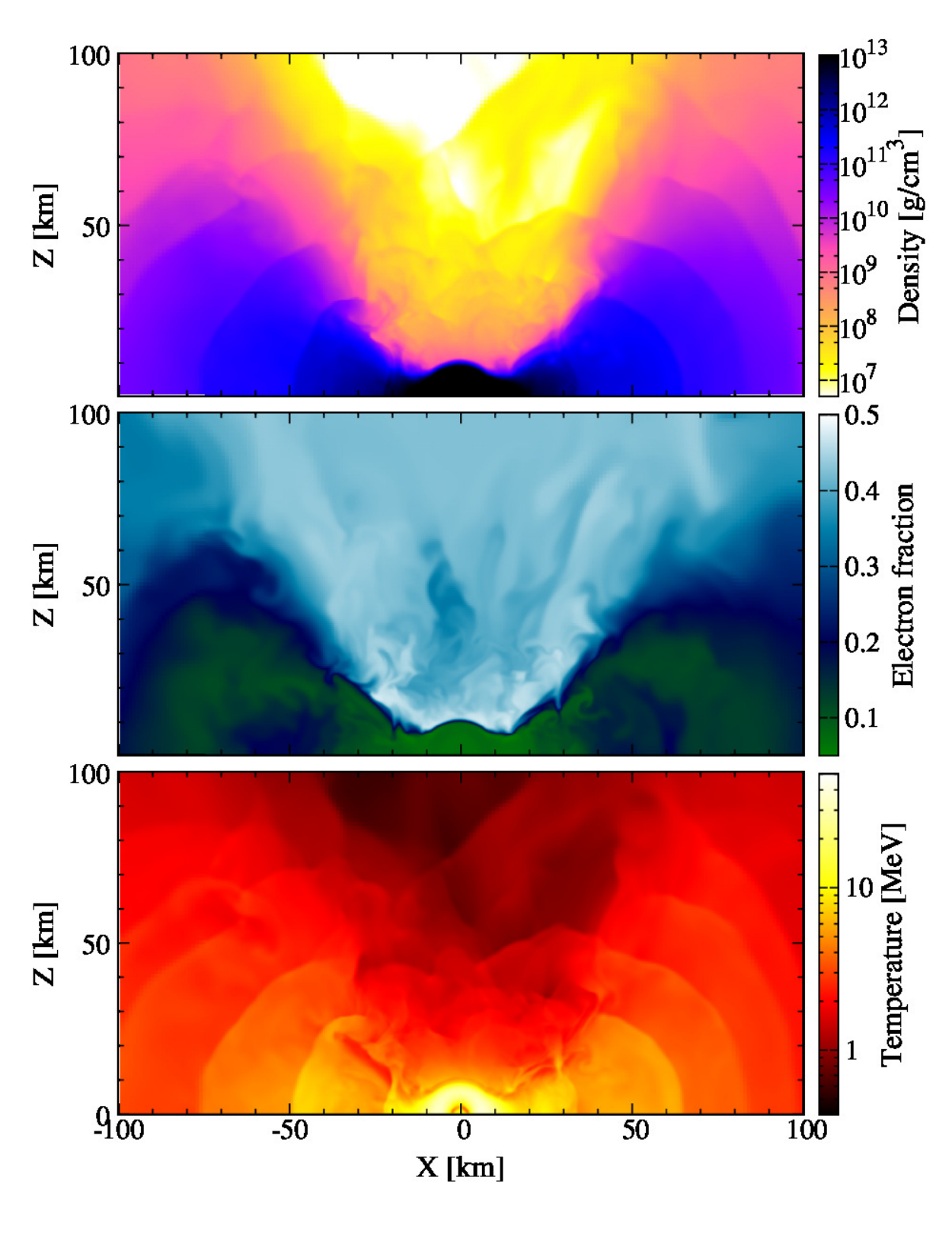}
\end{center}
\vspace{-3mm}
\caption{Profiles of the rest-mass density (top in each panel),
  electron number per baryon (middle in each panel), and temperature
  (bottom in each panel) in $x$-$z$ plane for SFHo-135-135h (top
  left), SFHo-125-145h (top right), DD2-135-135h (bottom left), and
  DD2-125-145h (bottom right) at 30\,ms after the onset of the merger.
  The filled circles (in black or white) in the top panels denote the
  inside of black holes.
\label{fig5}}
\end{figure*}

We briefly touch on the properties of the merger remnants located
around the central region because the torus around the central merger
remnant could be the origin of the further long-term mass ejection
(e.g., Refs.~\cite{Fernandez,Perego2014,Just2015}).  For the SFHo
models, the outcome for $t-t_{M_{-6}} \gtrsim 15$\,ms is a rotating black
hole surrounded by a massive torus irrespective of the mass ratio, as
displayed in Fig.~\ref{fig5}. For the SFHo-135-135 model, the torus
mass is $\approx 0.05M_\odot$ and its maximum density is less than
$10^{12}\,{\rm g/cm^3}$. For such relatively low density, the electron
degeneracy is not very high and also neutrinos escape efficiently from
the torus because the optical depth is small. 

On the other hand, for the SFHo-125-145 model (also for the
SFHo-130-140 model), the torus mass and maximum density are higher
than those for the SFHo-135-135 model.  In this case the maximum
density is higher than $\sim 10^{12}\,{\rm g/cm^3}$, the electron
degeneracy is higher than that for the SFHo-135-135 model, and a part
of neutrinos is trapped.  Then, the $\beta$-equilibrium among
neutrons, protons, electron, and neutrinos as $n + \nu_e
\leftrightarrow p + e^-$ and $p + \bar\nu_e \leftrightarrow n + e^+$
is approximately satisfied. Since the electron degeneracy is high, the
resulting value of $Y_e$ is lower than that for the SFHo-135-135
model.

Irrespective of the binary mass asymmetry, the resulting compact torus
has high temperature $\sim 10$\,MeV and is cooled dominantly by the
neutrino emission. Hence the torus is the neutrino-dominated accretion
torus. The order of magnitude for the neutrino luminosity (for $\nu_e$
and $\bar\nu_e$) is $10^{52}\,{\rm ergs/s}$ (see Fig.~\ref{fig3}).
Thus, the pair annihilation of neutrinos and anti-neutrinos to the
electron-positron pair, which is not taken into account in our present
simulation, would be activated and could modify the dynamics of the
merger remnants~(e.g., Refs.~\cite{ZB11,Just2016}).  In addition, the
system has a low density region above the black-hole pole. Such a system
satisfies the conditions for the central engine of short-hard gamma-ray
bursts.

Massive tori will be subsequently evolved by magnetohydrodynamics
(MHD) or viscous processes in reality: Angular momentum inside the
tori will be redistributed and associated with this effect, matter in
the tori will be heated up. Then, the geometrical thickness of the
tori will be increased, and possibly, an outflow that ejects the
matter from the outer part of the tori could be
launched~\cite{Fernandez,Perego2014,Just2015,kiuchi2014,martin}. The
total rest mass of the ejected matter could reach 10\% of the initial
torus mass, according to the previous studies. This suggests that the
ejecta with mass of the order $0.01M_\odot$ could follow the dynamical
mass ejection. We need to explore this process in our future study.
On the other hand, the luminosity of neutrinos emitted is not as high
as that by the remnant MNS. Thus, neutrino irradiation would not be as
important as the MHD/viscous effect for the mass ejection in the black
hole-torus system.

For the DD2 models, the final outcome is a MNS surrounded by a massive
torus as displayed in Fig.~\ref{fig5}. Although the central object is
different from a black hole, the surrounding matter distribution and
velocity profile (close to the Keplerian motion) are similar to those
for the SFHo models. Because the density of the MNS and torus is
higher than the torus surrounding the black hole found in the SFHo
models, the low value of $Y_e$ caused by the electron degeneracy is
clearly observed in the DD2 models. As in the torus surrounding black
holes, the torus around the MNS would be subject to the MHD or viscous
effects~\cite{martin}, and hence, it is natural to expect a
substantial fraction of mass ejection from the surrounding matter.
Because the MNS is long-lived for the DD2 models, it is also natural
to expect that the neutrino irradiation to the surrounding matter
plays an important role for inducing long-term mass ejection.

In the DD2 models, the torus mass and torus extent for the asymmetric
binaries are larger than that for the equal-mass one as in the SFHo
models.  This shows that the binary asymmetry increases not only the
dynamical ejecta mass but also the torus mass. This suggests that the
mass of the matter ejected by subsequent MHD/viscous effect would be
also enhanced in the asymmetric models.

The outer part of the torus surrounding the central object, that is
most subject to the mass ejection from the torus, is in general hot
and the value of $Y_e$ is not very small ($\agt 0.35$). This suggests
that the ejecta would not be very neutron-rich and less subject to
producing the heavy r-process elements, although they could be subject
to producing relatively light r-process elements. Exploring the 
torus-originated components of the ejecta in a self-consistent study
from the merger simulation throughout the subsequent remnant evolution
will be an important issue to fully understand the mass ejection
mechanism in the binary-neutron-star merger event. We plan to 
explore this issue in our future work. 

It is interesting to point out that for the DD2 models, the density in
the region above the MNS pole is as low as $\alt 10^7\,{\rm g/cm^3}$
for $t-t_{M_{-6}} \agt 20$\,ms. Since the luminosity of electron
neutrinos and anti-neutrinos emitted from the remnant MNS is high,
$\sim 10^{53}\,{\rm ergs/s}$, for the DD2 models, the $\nu_e
\bar\nu_e$ pair annihilation would be active near the MNS. According
to a simple order of magnitude estimate, the pair annihilation
luminosity is given by (e.g., Refs.~\cite{Cooperstein1987,ZB11})
\begin{eqnarray}
&&L_{\nu_e\bar\nu_e} \sim 10^{50}\,{\rm ergs/s}
\,\left({r \over 10^7\,{\rm cm}}\right)^{-1}
\left({\langle \epsilon_{\nu_e} \rangle + 
\langle \epsilon_{\bar\nu_e}\rangle \over 20\,{\rm MeV}}\right)
\nonumber \\
&&~~~~~
\times \left({L_{\nu_e} \over 10^{53}\,{\rm ergs/s}}\right)
\left({L_{\bar\nu_e} \over 10^{53}\,{\rm ergs/s}}\right) \nonumber \\
&&~~~~~
\times
\left({\cos\Theta \over 0.1}\right)^2 
\left({\theta_{\rm open} \over 0.1}\right)^{-2}, 
\end{eqnarray}
%%
%where $E_{\nu_e}$ and $E_{\bar\nu_e}$ the typical energy of $\nu_e$
%and $\bar\nu_e$, $L_{\nu_e}$ and $L_{\bar\nu_e}$ the luminosity of
%$\nu_e$ and $\bar\nu_e$, and 
where $\Theta$ is the typical angle of the collision between $\nu_e$
and $\bar\nu_e$, $r$ and $\theta_{\rm open}$ denote, respectively, the
extent and opening angle above the MNS pole, in which the pair
annihilation is enhanced.  This luminosity is high enough for
launching short-hard gamma-ray bursts like GRB\,130603B even for the
case that the merger remnant is surrounded by dynamical ejecta, as
demonstrated in~Ref.~\cite{Nagakura2014}.  Because the density of the
polar region in the vicinity of the MNS is low, high specific entropy
would be achieved in the presence of the $\nu_e \bar\nu_e$ pair
annihilation. This suggests that a strong outflow or a jet may be
launched from this system. If a sufficiently high specific entropy is
achieved, a relativistic jet responsible for a short-hard gamma-ray
burst could be indeed launched even from the remnant MNS. Including
the $\nu_e \bar\nu_e$ pair annihilation in our simulation will be an
important next step. 

\section{Summary and discussion}

We have reported our latest numerical results of neutrino radiation
hydrodynamics simulations for binary-neutron-star mergers in general
relativity, focusing on the dynamical mass ejection from the merger of
asymmetric binary neutron stars with typical mass for each neutron
star ($1.25$--$1.45M_\odot$) and with two representative
finite-temperature EOS. The following is the summary of our finding:
\begin{enumerate}
\item The dynamical ejecta mass depends weakly on the mass ratio for the
SFHo (soft-EOS) models. The reason for this is that while the
dynamical mass ejection from equal-mass or nearly equal-mass system is
induced primarily by shock heating and this effect becomes weak with
the increase of the degree of the binary asymmetry, the tidal effect
compensates the weakened shock-heating effect for the mass ejection in
the asymmetric systems.  
\item The dynamical ejecta mass depends significantly on the binary
  asymmetry degree for the DD2 (moderately stiff-EOS) models; it is
  $\approx 2 \times 10^{-3}M_\odot$ for the equal-mass case while it
  is $\approx 5 \times 10^{-3}M_\odot$ for the $1.25$--$1.45M_\odot$
  model. The reason for this is that the tidal torque, which plays a
  major role for the dynamical mass ejection in this EOS, is simply
  enhanced.
\item The averaged value of $Y_e$ decreases appreciably with the
  increase of the degree of the binary asymmetry irrespective of the
  EOS employed, and the peak value of $Y_e$ becomes less than 0.2 for
  the $1.25$--$1.45M_\odot$ models.
\item $Y_e$ of the ejecta has a broad mass distribution between
  $\approx 0.05$ and $\approx 0.5$ irrespective of the EOS and mass
  ratios. This property is well-suited for producing a variety of
  r-process heavy elements as illustrated in
  Refs.~\cite{Wanajo,radice2016}.
\item The neutrino irradiation effect to the dynamical ejecta, which
  is clearly found for the DD2 models, becomes weak as the binary
  asymmetry degree increases. The reason for this is that binary
  asymmetry reduces the shock heating efficiency at the onset of the
  merger, and as a result, the temperature of the remnant MNS is
  decreased, reducing the luminosity of the neutrino emission from the
  MNS.
\end{enumerate}

In our previous paper~\cite{Sekig2015}, we found for the equal-mass
binary merger that the total ejecta mass is larger for softer EOS. It
exceeds $0.01M_\odot$ only for the case that $R_{1.35} \alt 12$\,km
and it is of the order $10^{-3}M_\odot$ for $R_{1.35} > 13$\,km.  For
the case that the ejecta mass might be of the order $10^{-3}
M_{\odot}$, it would be too small to explain the total mass of
$r$-process heavy elements (the so-called second and third-peaks
elements) in our galaxy, unless the galactic merger rate of binary
neutron stars is unexpectedly high~\cite{hotoke2015} or some other
ejection mechanisms such as the disk wind are present.  Our present
simulations show that the ejecta mass can be increased in the presence
of an appreciable mass asymmetry of the binary systems even for the
case that $R_{1.35}=13.2$\,km.  This suggests that even if the EOS is
not very soft, the observed total mass of the r-process heavy elements
in our galaxy may be explained in the presence of a substantial
fraction of the asymmetric merger events. Here, we stress that even
from such asymmetric systems, neutron-rich matter with a variety of
$Y_e$ could be ejected.

Nevertheless, if a large fraction of the asymmetric binary merger
  has a mass ratio of $q\lesssim 0.9$, the averaged value of $Y_e$
would be small $\alt 0.2$ even if the EOS is soft. In such case,
although a substantial amount of the heavy r-process elements
around the second and third peaks could be produced, the light
  elements around the first peak would not be significantly
produced~\cite{Wanajo,radice2016}.  If this scenario is the case, we
have to rely on other components such as disk-wind
components~\cite{Fernandez,Just2015}, which can be produced in the
merger remnant for a time scale longer than the dynamical one.

As we mentioned above, the r-process elements are likely to be
produced in the neutron-rich ejecta. Because most of the produced
r-elements are unstable, they subsequently decay and the released
energy will be the source for an electromagnetic signal, in particular
in the near-infrared optical band~\cite{KN1,TH}.  Our present study
indicates that irrespective of the EOS and mass ratios, the ejecta
mass is larger than $10^{-3}M_\odot$. Under this condition, the
expected observed magnitude in the near-infrared optical bands is
smaller than 24~magnitude for an event at 100\,Mpc from the
earth. Such an event can be observed by Hyper-Suprime Cam (HSC) of the
Subaru telescope with one-minute-duration
observation~\cite{Tanaka}. Since HSC (in operation now) can
simultaneously observe a field of $\approx 1.75\,{\rm deg}^2$, a wide
field of $\sim 100\,{\rm deg}^2$ can be surveyed in one night by it.
Even if the position determination by gravitational-wave detectors is
not very good (e.g., Ref.~\cite{Nissanke}), this wide-field
observation will enable us to find a counterpart of the
gravitational-wave events.  These facts indicate that this
radio-actively powered electromagnetic signal is the promising
electromagnetic counterpart of binary-neutron-star mergers even for
the gravitational-wave observation with a small number of detectors
(by which the accuracy of the position determination is not very
high).

Light curves for this emission have been calculated for the dynamical
ejecta~\cite{KN1,TH}, based on the numerical results for it. 
Only in the presence of the dynamical ejecta, the luminosity
simply decreases with time in a power-law manner after the peak 
luminosity is reached in 1--10 days after the merger (the peak time
depends on the wave length). Here, in the presence of disk-wind
components, we will have two different types of the sources and hence
the electromagnetic signals from the ejecta will be significantly
modified~\cite{kasen15}.

For the observation of the electromagnetic counterparts, we need a
reliable theoretical prediction for the light curves. This is in
particular the case for searching the electromagnetic counterparts of
short duration. For this issue, we have to take into account all the
possible components other than the dynamical ejecta like the disk-wind
components.  We plan to explore this issue in the subsequent work.

\acknowledgments

We are grateful to M. Hempel for providing the EOS table data and to
M. Tanaka for helpful discussion on electromagnetic-counterpart
observation.  Numerical computations were performed on the
supercomputer K at AICS, XC30 at CfCA of NAOJ, FX10 at Information
Technology Center of Tokyo University, and SR16000 and XC30 at YITP of
Kyoto University.  This work was supported by Grant-in-Aid for
Scientific Research (24244028, 25103510, 25105508, 24740163, 26400267,
15K05077, 15H06857, 15H00783, 15H00836), for Scientific Research on
Innovative Area (24103001) of Japanese MEXT/JSPS, and by HPCI
Strategic Program of Japanese MEXT (project No. hpci140211 and
hpci150225). Kyutoku was supported by the RIKEN iTHES project.

%%%%%%%%%%%%%%%%%%%%%%%%%%%

%%%%%%%%%%%%%%%%%%%%%

\end{document}